\documentclass[12pt]{article}
\usepackage{jheppub}
\pdfoutput=1

\usepackage{amsmath,bbm,array,amsfonts,graphicx,wrapfig,lscape,float, mathtools,multirow,longtable,rotating,subfigure,ulem,cancel}

\newcommand{\be}{\begin{equation}}
\newcommand{\ee}{\end{equation}}
\newcommand{\beq}{\begin{equation}}
\newcommand{\beql}[1]{\begin{equation}\label{#1}}
\newcommand{\eeq}{\end{equation}}
\newcommand{\ba}{\begin{array}}
\newcommand{\ea}{\end{array}}
\newcommand{\bea}{\begin{eqnarray}}
\newcommand{\beal}[1]{\begin{eqnarray}\label{#1}}
\newcommand{\eea}{\end{eqnarray}}
\newcommand{\ben}{\begin{enumerate}}
\newcommand{\een}{\end{enumerate}}
\newcommand{\bean}{\begin{eqnarray*}}
\newcommand{\eean}{\end{eqnarray*}}
\newcommand{\eref}[1]{(\ref{#1})}
\newcommand{\sref}[1]{\S\ref{#1}}
\newcommand{\tref}[1]{Table~\ref{#1}}
\newcommand{\nn}{\nonumber}

\newcommand{\fref}[1]{Figure \ref{#1}}
\newcommand{\btab}[1]{\begin{tabular}{#1}}
\newcommand{\etab}{\end{tabular}}

\newcommand{\mesonic}{\mathcal{M}^{mes}}

\def \Black {\pdfsetcolor{0 0 0 1}}

\newcommand{\comment}[1]{}

\newcommand{\ud}{\mathrm{d}}

\newcommand{\PE}{\mathrm{PE}}
\newcommand{\PL}{\mathrm{PL}}

\newcommand{\qed}{\nobreak \ifvmode \relax \else
      \ifdim\lastskip<1.5em \hskip-\lastskip
      \hskip1.5em plus0em minus0.5em \fi \nobreak
      \vrule height0.75em width0.5em depth0.25em\fi}

\newcommand{\bC}{\mathbb{C}}

\newcommand{\bZ}{\mathbb{Z}}

\def\sl{\mathfrak{sl}}

\newcommand{\ft}[2]{{\textstyle\frac{#1}{#2}}}

\title{Mass-deformed Brane Tilings}

\author[a]{Massimo Bianchi,}
\author[b]{Stefano Cremonesi,}
\author[b]{Amihay Hanany,}
\author[a]{Jose Francisco Morales,}
\author[c]{Daniel Ricci Pacifici,}
\author[d]{Rak-Kyeong Seong}

\affiliation[a]{
{
Dipartimento di Fisica, Universit\`a di Roma ``TorVergata'' and I.N.F.N., Sezione di Roma ``TorVergata'', Via della Ricerca Scientifica, 00133 Roma, Italy
}
}

\affiliation[b]{
Theoretical Physics Group, The Blackett Laboratory,
Imperial College London, \\
Prince Consort Road, London SW7 2AZ, United Kingdom
}

\affiliation[c]{
Dipartimento di Fisica e Astronomia, Universit\`a degli Studi di Padova and I.N.F.N, Sezione di Padova, Via Marzolo 8, 35131, Padova, Italy
}

\affiliation[d]{
School of Physics, Korea Institute for Advanced Study,\\
85 Hoegi-ro, Seoul 130-722, South Korea
}

\emailAdd{
massimo.bianchi, morales@roma2.infn.it,\\
s.cremonesi,  a.hanany@imperial.ac.uk, \\
daniel.riccipacifici@pd.infn.it,
rkseong@kias.re.kr
}

\preprint{
{
\begin{flushright}
Imperial/TP/13/AH/04\\
KIAS-P14038
\end{flushright}
}
}

\abstract{
We study renormalization group flows among ${\cal N} = 1$ SCFTs realized on the worldvolume of D3-branes probing toric Calabi-Yau singularities, thus admitting a brane tiling description. The flows are triggered by masses for adjoint or vector-like pairs of bifundamentals and are generalizations of the Klebanov-Witten construction of the ${\cal N} = 1$ theory for the conifold starting from the ${\cal N} = 2$ theory for the $\bC^2/\bZ_2$ orbifold. In order to preserve the toric condition pairs of masses with opposite signs have to be switched on. We offer a geometric interpretation of the flows as complex deformations of the Calabi-Yau singularity preserving the toric condition. For orbifolds, we support this interpretation by an explicit string amplitude computation of the gauge invariant mass terms generated by imaginary self-dual 3-form fluxes in the twisted sector. In agreement with the holographic $a$-theorem, the volume of the Sasaki-Einstein 5-base of the Calabi-Yau cone always increases along the flow. 
}

\begin{document}

\maketitle

\section{Introduction \label{s1n1}}

A large class of  $\mathcal{N}=1$ superconformal gauge theories in four dimensions can be realized on the worldvolume of a stack of D3-branes probing toric Calabi-Yau threefold singularities in Type IIB string theory. The field content and interactions of such gauge theories are elegantly described by a bipartite graph on the $2$-torus,  known as a dimer model or brane tiling \cite{Hanany:1997tb,Hanany:2005ve,Franco:2005rj}. The faces, edges and nodes of the brane tiling are associated to $U(N)$ gauge group factors, bifundamental/adjoint matter and superpotential terms respectively.

 A simple class of toric CY singularities consists of orbifolds of the form $\mathbb{C}^3/\Gamma$ where $\Gamma$ is a finite Abelian subgroup of $SU(3)$ \cite{Davey:2010px,Davey:2011dd,Hanany:2010ne,Hanany:2011iw}. For Abelian $\Gamma$,  the brane tiling is made of $n=|\Gamma| $ hexagonal faces covering the torus. The superpotential, inherited from the parent $\mathcal{N}=4$  theory, is cubic in this case and the global symmetry is at least $U(1)^3$, including the $U(1)_R$ R-symmetry of the $\mathcal{N}=1$ superconformal theory. 
 
  When one of the complex planes is invariant under $\Gamma$, the singularity is actually of the form $\mathbb{C}^2/\mathbb{Z}_n \times \mathbb{C}$ and supersymmetry is enhanced to $\mathcal{N}=2$. These theories admit mass deformations which break ${\cal N}=2$ supersymmetry down to ${\cal N}=1$. Integrating out the massive fields, the theory flows to an $\mathcal{N}=1$ superconformal fixed-point in the IR with a quartic -- in general non-toric -- super-potential. As shown by Klebanov and Witten \cite{Klebanov:1998hh}, the conifold singularity  can be obtained in this way, by perturbing the quiver gauge theory on the worldvolume of D3-branes probing the $\mathbb{C}^2/\mathbb{Z}_2 \times \mathbb{C}$ orbifold singularity with mass terms for the adjoint fields. Toricity is preserved when the two mass parameters are equal and opposite. The brane tiling of the conifold theory is made of two square faces covering the torus like a chessboard \cite{Hanany:2005ve}. 

This construction admits a natural generalization to RG flows triggered by mass deformations of $\mathcal{N}=2$ superconformal theories for D3 branes probing $\mathbb{C}^2/\mathbb{Z}_n \times \mathbb{C}$ \cite{Gubser:1998ia}. The generalization starts with a quiver gauge theory with $n$ adjoint fields -- one for each $U(N)$ factor -- besides the bifundamental matter.  By giving masses to any of the adjoint fields, one generates a flow to a new -- in general non-toric -- ${\cal N}=1$ superconformal gauge theory. As we will review, toricity can be preserved by choosing the mass parameters to be all equal with alternating signs in a sequence of $k$ pairs of adjoints. In addition, a field redefinition is required in the IR to restore the toric property of the superpotential. The resulting quiver gauge theory is associated to the toric singularity $C(L^{k,n-k,k})$, where $C(X)$ denotes the cone over $X$, also known as a generalized conifold.\footnote{To simplify the notation, we drop $C(...)$ in naming Calabi-Yau cones and their corresponding brane tilings in the rest of this work. $L^{k,n-k,k}$ are members of a larger class of toric Sasaki-Einstein 5-fold called $L^{a,b,c}$ \cite{Cvetic:2005ft}, whose dual quiver gauge theories were found in \cite{Benvenuti:2005ja,Franco:2005sm,Butti:2005sw}.} 
Pictorially, the flow may be thought of as the result of squeezing $2k$ strips of hexagonal faces in the brane tiling down to $2k$ strips of squares. The case $\mathbb{C}^2/\mathbb{Z}_3 \times \mathbb{C}$ leading to the suspended pinch point (SPP) singularity \cite{Morrison:1998cs} -- also denoted by $L^{1,2,1}$ -- is an example which we consider in this work. 
 
The Type IIA dual description of these models {\`a} la Hanany-Witten, {\it i.e.} in terms of D4-branes wrapping a circle and suspended in between $n$ NS5-branes \cite{Uranga:1998vf,Dasgupta:1998su,vonUnge:1999hc}, suggests that mass deformations can be interpreted as complex deformations of the singularity. Indeed, when the $n$ NS5-branes are parallel to one another in the directions transverse to the D4-branes, the configuration enjoys $\mathcal{N}=2$ supersymmetry.  The adjoint chiral multiplet inside the $\mathcal{N}=2$ vector multiplet corresponds to moving the D4-branes along the NS5 brane or turning a Wilson line along the circle wrapped by the D4-branes. If the NS5-branes are rotated by generic angles, supersymmetry is completely broken. $\mathcal{N}=1$ supersymmetry is restored whenever the  NS5-branes wrap complex planes in $\mathbb{C}^3$. The masses of the adjoints are the complexified relative angles between NS5-branes.
The flow described above corresponds to the case where $k$ out of the $n$ NS5 branes are rotated. 

In the IIB description in terms of D3-branes on $\mathbb{C}^2/\mathbb{Z}_n \times \mathbb{C}$, the mass terms can be realized as imaginary self-dual fluxes of type (2,1)  as required by supersymmetry. Fluxes of this type arise from the twisted sector and  are associated to 3-forms $\omega_{(1,1)} \wedge dZ$ with $\omega_{(1,1)}$ being one of the twisted (1,1)-forms dual to an exceptional 2-cycle of the ALE hyperk\"ahler singularity $\mathbb{C}^2/\mathbb{Z}_n$ and $dZ$ being the holomorphic one-form on $\mathbb{C}$. Since one leg is non-compact, the flux is not quantized. While untwisted fluxes do not discriminate among the various nodes of the quiver, or the various faces of the brane tiling, twisted fluxes instead allow for different masses for the various fields as required for the RG flow. Moreover, the counting of twisted sectors, $n-1$ in this case, precisely matches the counting of independent mass deformations or the counting of relative rotations in the NS5-brane picture. Further evidence for this correspondence is obtained by computing the string disk amplitude involving the insertions of a 3-form field strength from the closed string twisted sector and two open string fermions. 
 
Our construction is not limited to flows starting from ${\cal N}=2$ orbifold quivers, but it applies to more general mass flows connecting two ${\cal N}=1$ superconformal theories. We consider cases where the starting theory is or is not of orbifold type. The flows in both cases are triggered by mass terms for bifundamental matter. As for deformations of ${\cal N}=2$ theories, we show that in order to restore toricity in the IR after integrating out the massive matter, one has to switch on equal mass parameters opposite in pairs, each giving mass to two chiral super-fields. 
We illustrate this construction for the flows $(\mathbb{C}^2/\mathbb{Z}_n \times \mathbb{C})/\mathbb{Z}_2 \to L^{k,n-k,k}/\mathbb{Z}_2$, $\mathbb{C}^3/\mathbb{Z}_{2n} \to L^{k,n-k,k}/\mathbb{Z}^\prime_2$ and $\mathrm{PdP}_{4b}\to \mathrm{PdP}_{4a}$.\footnote{Arising from D3-branes probing a non-compact CY 3-fold which is a complex cone over a toric (pseudo) del Pezzo surface $\mathrm{PdP}_{4}$.} A crucial r\^ole in these RG flows is played by accidental symmetries which appear after integrating out the massive fields and which restore the toric $U(1)^3$ symmetry in the IR. 

The plan of the paper is as follows. In section \sref{s1n1}, we review brane tilings and present the brane tiling for $\mathbb{C}^2/\mathbb{Z}_3 \times \mathbb{C}$ as an example. Section \sref{s2} then discusses mass deformations of brane tilings.
The focus is on the RG flow $\mathbb{C}^2/\mathbb{Z}_3 \times \mathbb{C} \to L^{1,2,1}$ and then on the generalizations to other RG flows: $(\mathbb{C}^2/\mathbb{Z}_n \times \mathbb{C})/\mathbb{Z}_2 \to L^{k,n-k,k}/\mathbb{Z}_2$, $\mathbb{C}^3/\mathbb{Z}_{2n} \to L^{k,n-k,k}/\mathbb{Z}^\prime_2$ and $\mathrm{PdP}_{4b} \to \mathrm{PdP}_{4a}$. Section \sref{srch} then discusses a-maximization and volume minimization in order to identify the superconformal R-symmetry at the IR fixed point and to check that central charges decrease along the RG flows in accordance with the $a$-theorem. Section \sref{sflow} discusses the complex structure deformations of the Calabi-Yau cones under mass deformations and the relation between Hilbert series of the toric Calabi-Yau cones associated to the UV and the IR theories. Section \sref{samp} outlines with an explicit computation of the disk amplitudes the correspondence between mass deformations in the boundary SCFT and the effect of 3-form fluxes in the bulk. The paper closes with concluding remarks in section \sref{sec:conclusions}. The appendix contains details of the RG flows induced by mass deformations; this includes the UV superpotential, the mass terms, the IR superpotential, and the field redefinitions, if any, which are needed to restore toricity. 
\\

\section{Brane tilings \label{s1n1}}

In this section we review brane tilings, which encode the data defining superconformal quiver gauge theories on the worldvolume of D3-branes at toric Calabi-Yau cone singularities. Brane tilings lead to a powerful \textit{forward} algorithm \cite{Feng:2000mi,Feng:2001xr,Hanany:2005ve,Franco:2005rj,Franco:2006gc} for computing the vacuum moduli spaces of superconformal quiver gauge theories.
\\
    
\paragraph{Brane tiling dictionary.} A brane tiling is a bipartite graph on a 2-torus, \textit{i.e.} a covering of the 2-torus by even-sided polygonal faces bounded by edges connecting a black to a white node. The black or white coloring of nodes corresponding to the bipartiteness of the graph determines an orientation around vertices, which we take by convention to be clockwise around white nodes and counter-clockwise around black nodes. This in turn induces an orientation of the dual graph, the periodic quiver diagram of the gauge theory.

The brane tiling/gauge theory dictionary is as follows:
\begin{itemize}
\item \textbf{Faces} are associated to $U(N)_i$ gauge group factors. We use $F$ for the total number of faces.

\item \textbf{Edges} adjacent to faces $i$ and $j$ represent chiral superfields $X_{i\hspace{0.04cm}j}$ transforming in the bifundamental representation of the associated gauge groups $U(N)_i\times U(N)_j$. 
The quiver orientation of the bifundamental field $X_{i\hspace{0.04cm}j}$ is given by the orientation around the black and white nodes at the two ends of the corresponding tiling edge.
We denote by $E$ the total number of edges.

\item \textbf{White (black) nodes} correspond to positive (negative) monomial terms in the superpotential $W$ made of the products of the bifundamental fields associated to the edges which end on the node  and which are ordered in a clockwise (counter-clockwise) fashion. The bipartite nature of the graph implies that every field appears in the superpotential precisely once in a positive and once in a negative term.
We call this property the \textbf{toric condition} \cite{Feng:2000mi}.  
We denote by $V$ the total number of vertices.
\\
\end{itemize}

\paragraph{The incidence matrix  $d_{G\times E}$} incorporates the charges of the chiral fields under the $U(1)_i$ factor of the $U(N)_i= U(1)_i \times SU(N)_i$ gauge groups. $G$ is the number of gauge groups, which equals the number of faces $F$ of the tiling. 
In this work, we will concentrate on the Abelian case with $N=1$, such that the incidence matrix fully incorporates the gauge charges of the theory. The $i^{\rm th}$-entry is  $-1$ for $X_{i\hspace{0.04cm}j}$, $+1$ for $X_{j\hspace{0.04cm}i}$ and zero otherwise.
The matrix $d_{G\times E}$ has $G-1$ linearly independent rows that can be collected in a separate matrix $\Delta_{G-1\times E}$. 
\\

\paragraph{The Kasteleyn matrix K} is a matrix which encodes information about the connectivity of the bipartite diagram.  Rows and columns index black nodes $b_m$ and white $w_n$ nodes respectively and the entries of the matrix are associated to edges. An edge $X(m,n)$ between nodes $(b_m,w_n)$ has a winding number $(h_a,h_b)$ associated to the $a,b$-cycles and the boundaries of the fundamental domain of the 2-torus. Accordingly, elements of the Kasteleyn matrix take the following form,
\beal{esk}
K_{mn} (z_1, z_2) = \sum_{X(m,n)} z_1^{h_a(X(m,n))} z_2^{h_b(X(m,n))} ~,~ 
\eea
where $z_1,z_2$ are the fugacities for the winding numbers along the $a$- and $b$-cycles of the 2-torus respectively.

The permanent of the Kasteleyn matrix, also known as the characteristic polynomial of the brane tiling,
\be
\text{perm}~ K(z_1, z_2)=\sum_{n_i}  c_{n_1,n_2} z_1^{n_1} z_2^{n_2}
\ee
encodes the toric diagram of the singularity. More precisely, the absolute values of the coefficients $|c_{n_1,n_2}|$ give the multiplicities of the points $(n_1,n_2)$ in the toric diagram.
\\

\paragraph{Perfect matching.} A perfect matching \cite{Franco:2005rj,Franco:2006gc} is a collection $p_\alpha$ of edges in the brane tiling which includes every white and black node precisely once. 
Each perfect matching contributes a monomial to the determinant of the Kasteleyn matrix and vice versa. Perfect matchings therefore can be associated to points in the toric diagram of the Calabi-Yau 3-fold. We denote by $c$ the total number of perfect matchings.
 
Perfect matchings are summarized in a perfect matching matrix $P_{E\times c}$ with entries
\beal{e200n0}
P_{\ell \alpha} = \begin{cases}
1 & ~~\text{if}~ X_\ell \in p_\alpha\nn\\
0 & ~~\text{if}~ X_\ell \notin p_\alpha
\end{cases}
~~.
\eea
To each perfect matching we associate a perfect matching variable which we denote by the same letter $p_\alpha$ with a slight abuse of notation. Perfect matching variables can be interpreted as fields in a gauged linear sigma model (GLSM) without superpotential, whose moduli space is the toric Calabi-Yau singularity. 
Indeed, each bifundamental field $X_\ell$ in the Abelian toric quiver gauge theory can be expressed as a product of perfect matching variables $p_\alpha$ via the simple relation
\beal{e200n1}
X_\ell = \prod_{\alpha} (p_\alpha)^{P_{\ell\alpha}}
\eea
in such a way that the F-term constraints of the toric quiver gauge theory are automatically satisfied.
\\

\paragraph{Zig-zag paths.} One can identify a particular set of paths along edges of the brane tiling which are known as zig-zag paths \cite{Hanany:2005ss}. A zig-zag path $\eta_i$ is a closed non-trivial path on the 2-torus along the edges of the brane tiling. The edges are selected in such a way that the path makes a maximal left turn on a white node and a maximal right turn on a black node. As an oriented path on the torus, each zig-zag path has winding numbers along the $a$- and $b$-cycles of the torus. The winding numbers encode a fan in the plane which relates to a $(p,q)$-web diagram \cite{Aharony:1997bh}. The dual of the web diagram is the toric diagram \cite{Feng:2005gw}.   
A bipartite graph on a torus which has zig-zag paths that do not self-intersect is considered to be consistent and to realize a unitary superconformal quiver gauge theory \cite{2011arXiv1104.1592B,2009arXiv0901.4662B}. 
\\

\paragraph{Mesonic moduli space.} The vacuum moduli space  resulting from imposing both F- and D-terms constraints of the $4d$ $\mathcal{N}=1$ supersymmetric Abelian gauge theory is  a non-compact toric Calabi-Yau threefold. 
Using the basis of GLSM fields represented by perfect matching variables, the constraints can be obtained as follows \cite{Feng:2000mi, Feng:2001xr, Feng:2002zw,Feng:2002fv,Hanany:2005ve,Franco:2005rj,Franco:2006gc}.
\begin{itemize}
\item The \textbf{F-terms} $\partial_X W =0$ are solved thanks to the introduction of perfect matching variables, which are defined modulo an Abelian gauge symmetry which leaves \eqref{e200n1} invariant. The charges of the perfect matching variables under this gauge symmetry are encoded in the charge matrix
\bea
Q_{F~(c-G-2)\times c} = \ker{(P_{E\times c})}~~,
\eea
 
\item The \textbf{D-term} charge matrix $Q_{D~(G-1)\times c}$ is defined by the relation
\beal{e200n21}
\Delta_{(G-1)\times e}=Q_{D~(G-1)\times c}{\cdot}P_{c\times e}^{t}~~.
\eea
where $\Delta_{(G-1)\times e}$ is the matrix formed by the $G-1$ independent rows of $d_{G\times e}$.\footnote{The number of incoming and outgoing arrows at each quiver node is the same, ensuring gauge anomaly cancellation. This results in $\Delta_{(G-1)\times e}$ which forms the $G-1$ independent rows of the quiver incidence matrix $d_{G\times e}$.}

\item One can combine the $F$- and $D$- charges into the total charge matrix of the GLSM,
\beal{e200n22}
Q_{t~(c-3)\times c} =
\left(
\ba{c}
Q_F \\ Q_D
\ea
\right)~~.
\eea

\end{itemize}

The mesonic moduli space of the Abelian toric quiver gauge theory can be expressed as a K\"ahler quotient of the ring of perfect matching variables $\mathbb{C}^c$ by the
$U(1)^{c-3}$-action $Q_t$, 
\beal{e200n23}
\mesonic = \mathbb{C}^c// Q_t~~.
\eea
The integer kernel of $Q_t$,
\beal{e200n23}
G_t = \ker(Q_t)~~,
\eea
is a matrix whose rows are the coordinates of the points in the toric diagram associated to the $c$ perfect matchings.
\\

\paragraph{The Hilbert series.} The Hilbert series is a generating function which counts chiral gauge invariant operators. The Hilbert series of the mesonic moduli space \eref{e200n23} counts the number of $U(1)^{c-3}$-invariant monomials made out of the $c$
perfect matching variables $p_\alpha$.  It is computed using the {\it Molien integral} 
\beal{e200n30}
g(t_\alpha;\mesonic) =
\prod_{i=1}^{c-3} \oint_{|z_i|=1} \frac{\ud z_i}{2\pi i z_i}
\prod_{\alpha=1}^{c}
\frac{1}{1-t_\alpha \prod_{j=1}^{c-3} z_j^{(Q_{t})_{j\alpha}}}
~~,
\eea
where $t_\alpha$ is the fugacity corresponding to the perfect matching variable $p_\alpha$.

The Hilbert series encodes information about the generators of the moduli space as well as the relations formed amongst them. This information can be extracted from the  so called {\it  plethystic logarithm} \cite{Benvenuti:2006qr,Feng:2007ur} of the Hilbert series $g_1(y_\alpha)$ defined as 
\beal{e200n31}
\PL[g(t_\alpha;\mesonic)]=\sum_{k=1}^{\infty} \frac{\mu(k)}{k} \log\left[ g(t_\alpha^k) \right]=\sum_i n_i  M_i(t_\alpha) \label{pl}
\eea
where $\mu$ is the M\"obius function, $M_i(t_\alpha)$ are monomials made of fugacities $t_\alpha$ and $n_i$ are integers. The Hilbert series can be  reconstructed from its plethystic logarithm and written in the simple product form 
 \be
 g(t_\alpha;\mesonic)  =\prod_{i} {1\over (1-M_i)^{n_i}} = \PE \Big[ \sum_i n_i M_i \Big]~,
 \ee
where $\PE$ refers to the \textit{plethystic exponential}.\footnote{The \textit{plethystic exponential} of a multivariate function $f(t_1, . . . , t_n)$ that vanishes at the origin, $f(0,...,0) = 0$, is defined as ${\rm PE} \left[ f(t_1, t_2, \ldots, t_n) \right] = \exp \left( \sum_{k=1}^\infty \frac{1}{k} f(t_1^k, \cdots, t_n^k) \right)$. Its inverse is the plethystic logarithm $\PL$.}
    In particular, when $\PL[g(t_\alpha;\mesonic)]$ contains a finite number of terms, the space $\mesonic$ is said to be a {\it complete intersection}.\footnote{When a complete intersection has only one relation, the corresponding space is called a hypersurface.} It is parametrized by the generators corresponding to the monomials $\{ M_i(t_\alpha) ; n_i>0\}$ satisfying a finite number of relations corresponding to the monomials $\{ M_i(t_\alpha) ; n_i<0\}$. 
    
Simple representatives in the class of complete intersections are Abelian orbifolds of the form $\mathbb{C}^2/\mathbb{Z}_n \times \mathbb{C}$ with the Hilbert series \cite{Benvenuti:2006qr}
 \beal{gznz2}
  g(t_\alpha; \mathbb{C}^2/\mathbb{Z}_n \times \mathbb{C} ) ={1\over n } \sum_{h=1}^n {1\over \prod_{\alpha=1}^3 (1- \omega_n^{a_\alpha h  } t_\alpha) } ={ (1- t_1^n t_2^n) \over (1-t_1^n )(1-t_2^n) (1-t_1 t_2) (1-t_3) }    
  \nn\\
 \eea
 with $\omega_n=e^{2\pi i \over n}$ and $a_\alpha=(1,-1,0)$. The result in the right hand side shows that the orbifold can be viewed
as a hypersurface $xy=w^n$ in $\mathbb{C}^4$ with coordinates $(x,y,w,z)$. 
 \\

\subsection{Example: $\mathbb{C}^2/\mathbb{Z}_3\times\mathbb{C}$ }

\begin{figure}[ht!!]
\begin{center}
  \begin{tabular}{ccc}
\includegraphics[trim=0cm 0cm 0cm 0cm,width=0.25\textwidth]{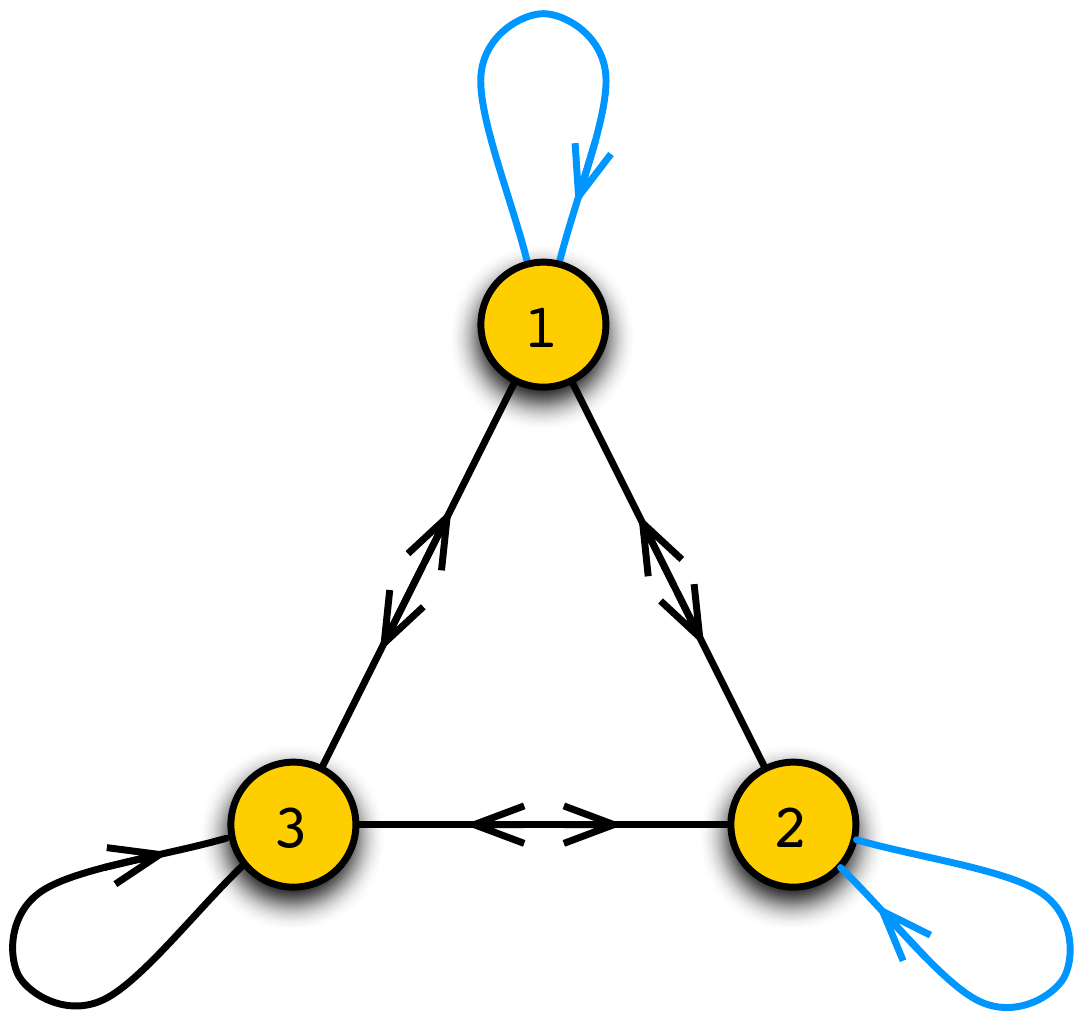}&~~
\includegraphics[trim=0cm 0cm 0cm 0cm,width=0.25\textwidth]{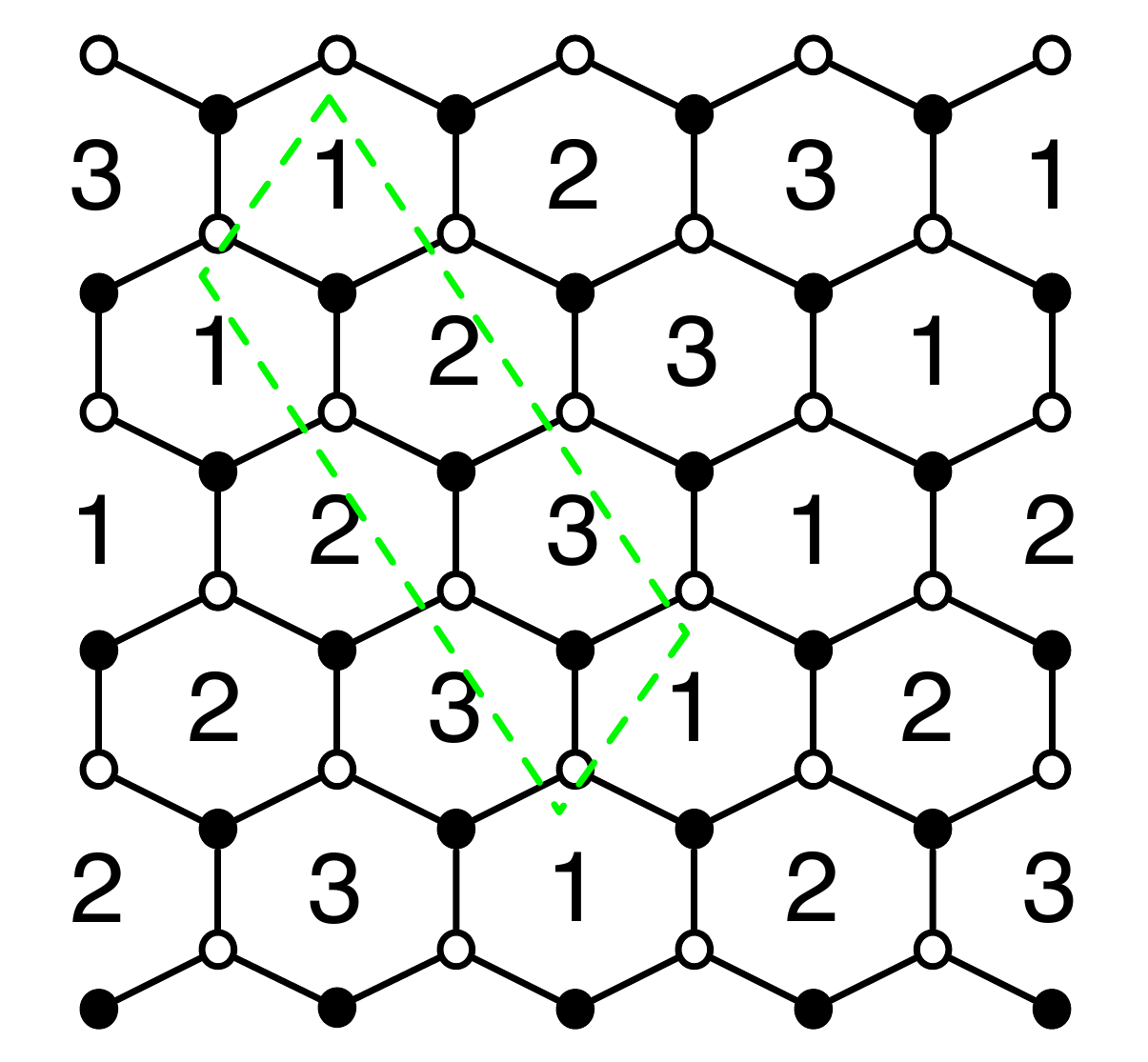}&~~ 
\includegraphics[trim=0cm 0cm 0cm 0cm,width=0.25\textwidth]{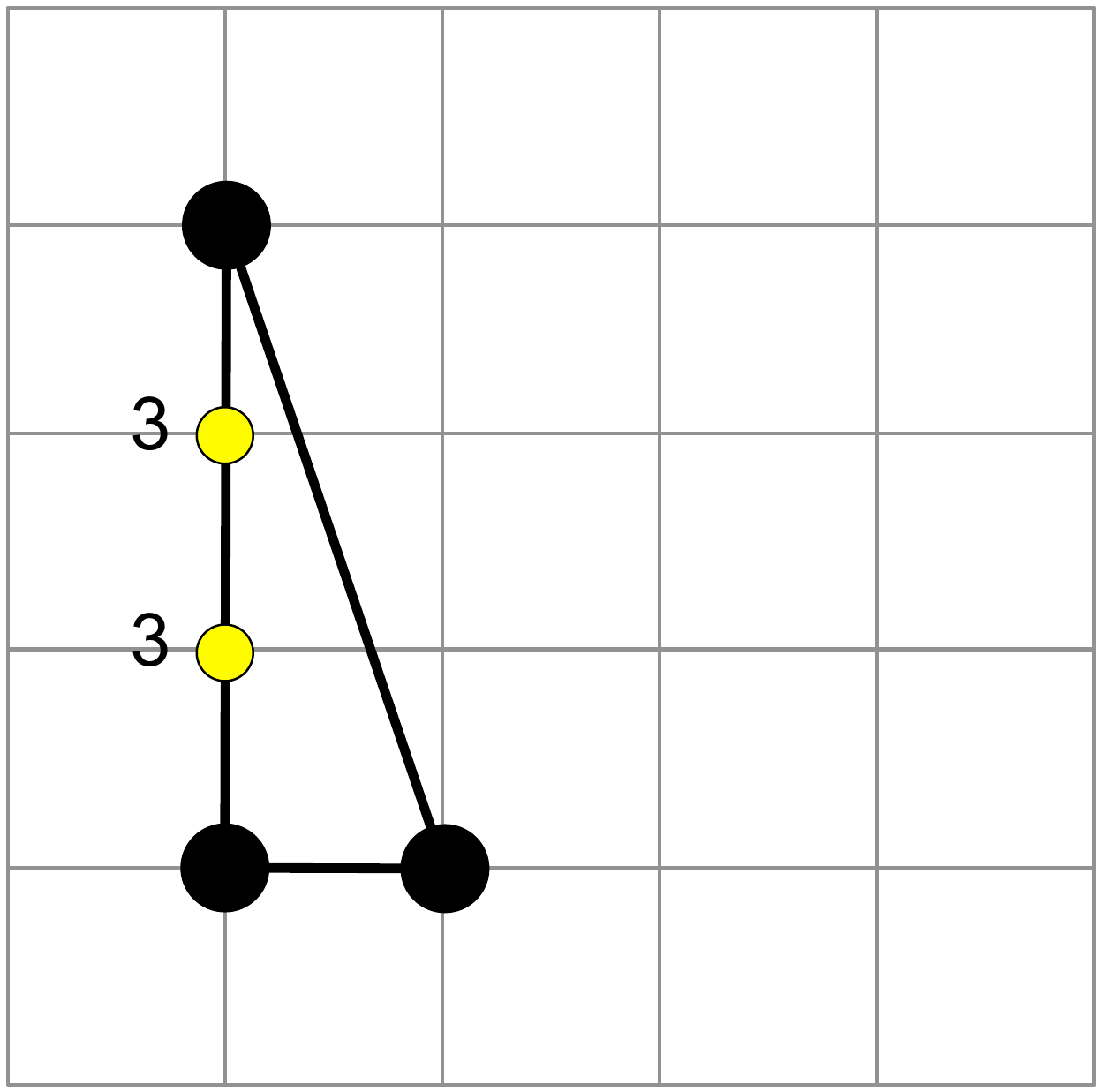} 
  \end{tabular}
\caption{
Quiver, dimer and toric diagrams for  $\mathbb{C}^2/\mathbb{Z}_3\times\mathbb{C}$  singularity.
\label{figc2z3}
}
 \end{center}
 \end{figure}

 Let us illustrate the brane tiling tools in the simple case of  $\mathbb{C}^2/\mathbb{Z}_3\times\mathbb{C}$.  The associated quiver, brane tiling and toric diagrams are displayed in \fref{figc2z3}.

The tiling is made of hexagons (as is always the case for orbifolds of $\mathbb{C}^3$) with $F=3$ faces associated to three $U(1)$ nodes in the quiver diagram and $E=9$ edges $X_{i\hspace{0.04cm}j}$, $i=1,2,3$ associated to chiral fields in the adjoint for $i=j$ and bifundamental representation otherwise.  
The Kasteleyn matrix is given by 
\be
K(z_1, z_2)=
\left(
\begin{array}{ccc}
 1+z_2 & 0  & z_1  \\
 1 & 1+z_2  & 0  \\
 0 & 1  & 1+z_2  
\end{array}
\right)
\ee   
  Each row is associated to a black node. For instance, there are three lines starting from the first black node (first row): two 
 towards the first white node contributing $1+z_2$ (one internal and one crossing the second boundary) and one towards 
 the third white node contributing $z_1$. This determines the first line of $K$.  The determinant is given by
  \be
\det K=1+z_1+3 z_2+3 z_2^2+z_2^3  \label{detk} 
\ee
Each monomial term in \eref{detk} is associated to a point in the toric diagram with coordinates given by the exponents of $z_{1,2}$. 
\\

\section{Mass deformations \label{s2}}

We are interested in mass deformations of brane tilings. Mass deformation requires the presence of adjoint fields $\phi_i=X_{i\hspace{0.04cm}i}$ or pairs of bifundamentals $(X_{i\hspace{0.04cm}j},X_{j\hspace{0.04cm}i})$ for a given ``$i$" different from ``$j$". The toric superpotential is deformed by adding quadratic mass terms
which violate the toric condition on the superpotential \cite{Feng:2000mi} and break the $U(1)^3$ mesonic and R- symmetry \cite{Butti:2006nk}. 
We are interested in the low energy theory that is obtained once the massive fields $\phi_i$ (or $X_{ij},X_{ji}$) have been integrated out. This low energy theory has a superpotential which generically violates the toric condition. In the following we will show that, for a large class of tilings and suitable mass deformations, the low energy theory has accidental symmetries which restore the toric $U(1)^3$ symmetry group. The accidental symmetry is made manifest by certain field redefinitions of the massless fields which restore the toric condition for the superpotential.

 We focus on brane tilings which have at least two adjacent strips of hexagonal faces and consider the effect of give mass to the hypermultiplets along these strips. 
 A strip made of a single type of faces leads to adjoint edges while two alternating faces along the strip leads to bifundamental matter. We collectively label the massive fields $X^{(m)}$ and the light fields $X^{(l)}$ to highlight their different r\^oles. We consider the mass-deformed superpotential
 \be
W_{\rm deformed}(X^{(m)},X^{(l)})=W(X^{(m)},X^{(l)})+\Delta W(X^{(m)})
 \ee
 where $W(X^{(m)},X^{(l)})$ is the initial toric superpotential and $\Delta W(X^{(m)})$ is deformation of one of the following types
 \begin{itemize}
\item \textbf{Adjoint:}\hspace{1.5cm} $\Delta W = \frac{m}{2} \, (\phi_{i_1}^2-\phi_{i_2}^2) $ ~~~~~\textit{or}
\item \textbf{Bifundamental:} $\Delta W =  m ( X_{i_1 j_1} X_{j_1 i_1} - X_{i_2 j_2} X_{j_2 i_2} ) $, 
\end{itemize}
possibly involving several pairs of mass terms, all with the same mass parameter $m$.

Integrating out the massive fields $X^{(m)}$, by solving their F-term equations  
 \be
\frac{\partial}{\partial X^{(m)}} W_{\rm deformed}(X^{(m)},X^{(l)})= 0
 \ee
in terms of the light components $X^{(l)}$, one finds a non-toric superpotential $W_{\rm low}(X^{(l)})$ for the light fields. We are able to restore the toric condition by field redefinitions of the light fields
  \beal{e200n40}
X_{i\hspace{0.04cm}j}^\prime = X_{i\hspace{0.04cm}j} + \frac{1}{m} \sum_{k} c^{(ij)}_{k} X_{i\hspace{0.04cm}k} X_{k\hspace{0.04cm}j}  \qquad {\rm or} \qquad \phi_i^\prime = \phi_i + \frac{1}{m}\sum_{k} c^{(ij)}_{k} X_{i\hspace{0.04cm}k} X_{k\hspace{0.04cm}j} 
\eea
with some judicious choice of the coefficients $c^{(ij)}_{k}$ .
The low energy superpotential $W_{\rm low}(X^{(l)})$, rewritten in terms of the new variables $X^{(l)}{}'$  can be shown to satisfy the toric condition. As such the IR fixed point of the RG flow is associated to a new brane tiling. In the following sections we consider some simple examples of mass-deformed brane tilings.
 \\

 \begin{figure}[h!]
\begin{center}
      \includegraphics[trim=0cm 0cm 0cm 0cm,width=1\textwidth]{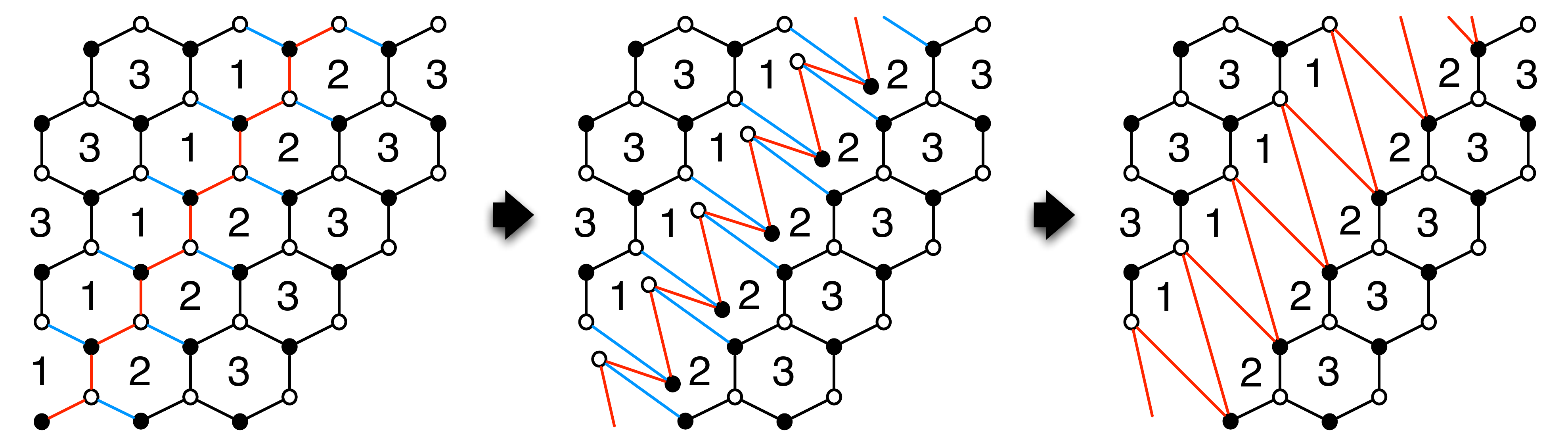}
       \end{center}
 \caption{Flow from $\mathbb{C}^2/\mathbb{Z}_3 \times \mathbb{C}$ to $L^{1,2,1}$. }
 \label{figstripmass}
 \end{figure}   
 
 \begin{figure}[h!]
\begin{center}
 \includegraphics[trim=0cm 0cm 0cm 0cm,width=0.7\textwidth]{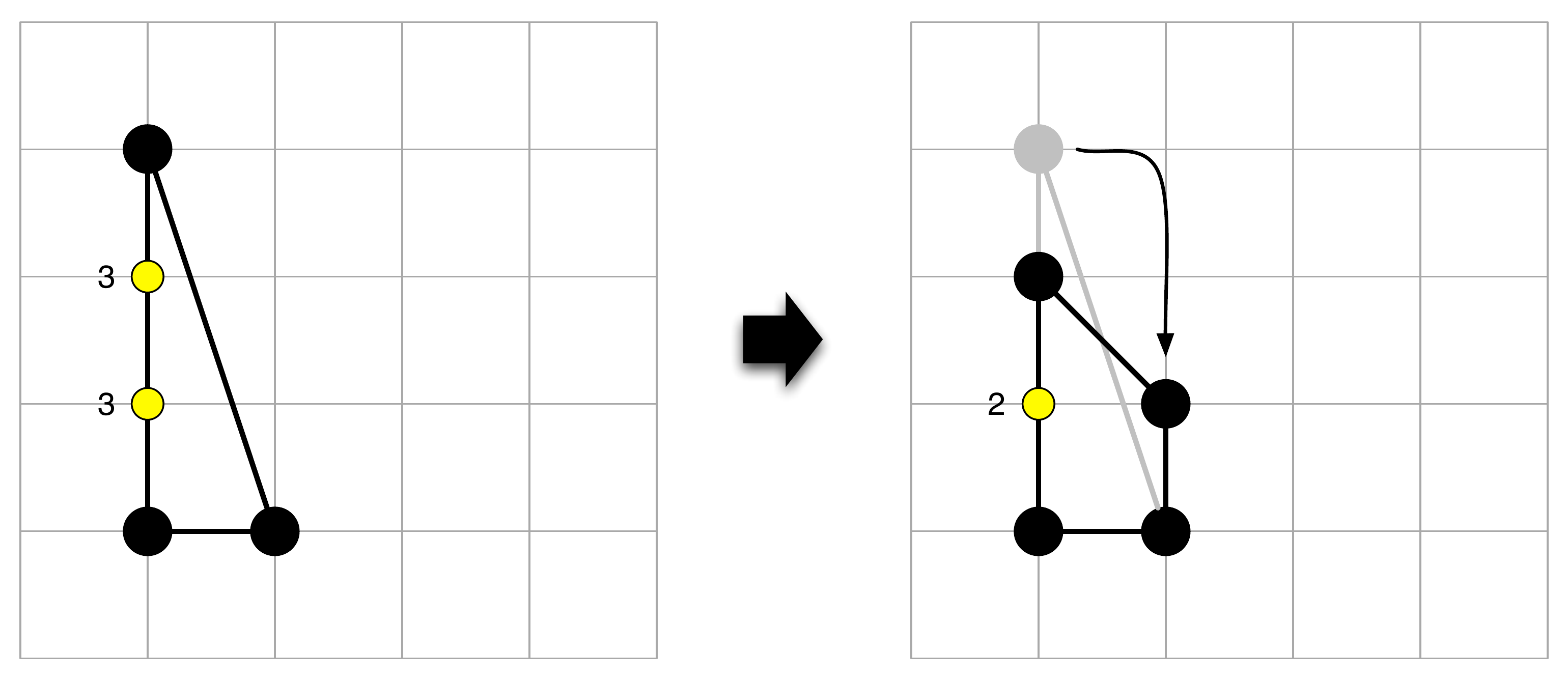} 
\caption{
Toric diagrams for the $\mathbb{C}^2/\mathbb{Z}_3\times\mathbb{C}$ case (on the left) and for SPP (on the right) before and after the mass deformation. Mass deformation amounts to a displacement of an extremal toric point. \label{Z3case}}
 \end{center}
 \end{figure}
 
\subsection{$\mathbb{C}^2/\mathbb{Z}_3 \times \mathbb{C}$  to SPP ($L^{1,2,1}$)}

We first illustrate mass deformation of brane tilings in our working example of  $\mathbb{C}^2/\mathbb{Z}_3 \times \mathbb{C}$. The starting super-potential is
\beal{es95n1}
W= \phi_1 \left( X_{1\hspace{0.04cm}3} X_{3\hspace{0.04cm}1} -X_{1\hspace{0.04cm}2} X_{2\hspace{0.04cm}1}\right)
+ \phi_2 \left(X_{2\hspace{0.04cm}1} X_{1\hspace{0.04cm}2}- X_{2\hspace{0.04cm}3} X_{3\hspace{0.04cm}2}  \right)
+ \phi_3 \left( X_{3\hspace{0.04cm}2} X_{2\hspace{0.04cm}3} -X_{3\hspace{0.04cm}1} X_{1\hspace{0.04cm}3} \right) ~~
\eea
The corresponding brane tiling and quiver diagram are shown in \fref{figc2z3}. We consider the mass deformation
 \beal{es95n2}
\Delta W =  \frac{m}{2} \left(\phi_1^2 - \phi_2^2\right)~~.
\eea
In the Type IIA  description, this corresponds to rotating the NS5-brane  between D4-branes 1 and 2.

 \begin{figure}[t!]
\begin{center}
 \includegraphics[trim=0cm 0cm 0cm 0cm,width=0.9\textwidth]{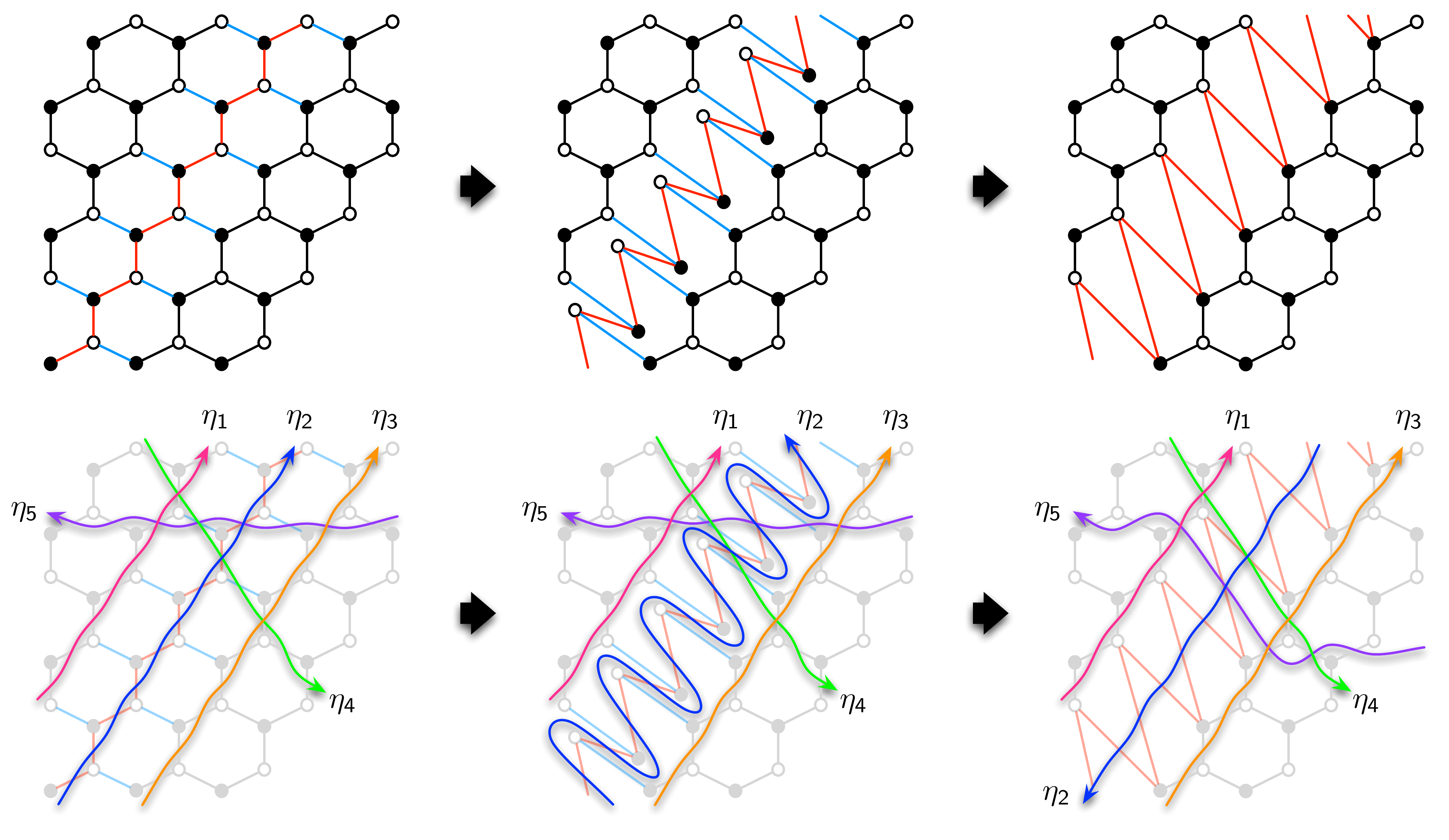} 
\caption{Mass deformation for the brane tiling of $\mathbb{C}^2/\mathbb{Z}_3\times \mathbb{C}$ with zig-zag paths. $\eta_2$ inverts its direction during the mass deformation while the adjacency of the zig-zag paths is preserved.  \label{zigzagdef}
}
 \end{center}
 \end{figure}

 Along the flow, massive fields can be integrated out. The effective superpotential at the infrared end of the flow is found by solving
   the mass-deformed F-term conditions for  $\phi_1$ and $\phi_2$  in favor of the light components   
\beal{es95n3}
\phi_1& = &\frac{1}{m} \left(X_{1\hspace{0.04cm}2} X_{2\hspace{0.04cm}1} - X_{1\hspace{0.04cm}3} X_{3\hspace{0.04cm}1} \right)~~,~~
\phi_2 = \frac{1}{m} \left(X_{2\hspace{0.04cm}1} X_{1\hspace{0.04cm}2} - X_{2\hspace{0.04cm}3} X_{3\hspace{0.04cm}2} \right)~~.
\eea
 Substituting this into the deformed superpotential $W_{\rm deformed}=W+\Delta W$ one finds a non-toric superpotential. Remarkably, a toric superpotential is recovered under the field redefinitions
\beal{es95n4}
\phi_3 &=& \phi_3^\prime - \frac{1}{2m} \left(X_{3\hspace{0.04cm}1} X_{1\hspace{0.04cm}3} + X_{3\hspace{0.04cm}2} X_{2\hspace{0.04cm}3} \right)
~~\nn\\
X_{1\hspace{0.04cm}2}X_{2\hspace{0.04cm}1} &=& m X_{1\hspace{0.04cm}2}^\prime X_{2\hspace{0.04cm}1}^\prime ~~, 
\eea
Indeed, by plugging \eref{es95n3} and \eref{es95n4} into $W_{\rm deformed}$, one finds the superpotential
\beal{es95n5}
W_{\rm final} = \phi_3 \left(X_{3\hspace{0.04cm}2} X_{2\hspace{0.04cm}3}-X_{3\hspace{0.04cm}1} X_{13}\right)
+ X_{1\hspace{0.04cm}2}^\prime X_{2\hspace{0.04cm}1}^\prime X_{1\hspace{0.04cm}3} X_{3\hspace{0.04cm}1}- X_{2\hspace{0.04cm}1}^\prime X_{1\hspace{0.04cm}2}^\prime X_{2\hspace{0.04cm}3} X_{3\hspace{0.04cm}2} ~~.
\eea
The field content and the superpotential in \eref{es95n5} are those of the suspended pinch point theory \cite{Feng:2000mi,Feng:2002fv,Hanany:2005hq}, also known as the theory for $L^{1,2,1}$. 

Interestingly, the result of integrating out the massive fields and shifting the light fields can be visualized in the brane tiling as squeezing two consecutive strips of hexagonal faces into two strips of square faces as illustrated in \fref{figstripmass}. The left side of \fref{figstripmass} shows the brane tiling for $\mathbb{C}^2/\mathbb{Z}_3 \times \mathbb{C}$. Edges associated to massive fields are drawn in blue. These fields are aligned along a \textit{deformation strip} made of two neighboring arrays of hexagonal faces. The remaining internal edges in the deformation strip are drawn in red and form a closed cycle on the 2-torus which we call the \textit{deformation line}. In order to obtain the brane tiling of the low energy theory, we need to remove the blue edges and move the deformation line in such a way that white nodes and black nodes move towards the opposite boundaries of the closed deformation strip, as shown in the middle column of \fref{figstripmass}. There are two options for the choice of merging nodes of the same color, but in the current example they lead to equivalent theories. In more involved examples, like for the brane tiling of $\mathrm{PdP}_{4b}$ and its mass flow, only one of the two choices leads to a consistent tiling as illustrated in \fref{fflow71}.

 \begin{figure}[t!]
\begin{center}
 \includegraphics[trim=0cm 0cm 0cm 0cm,width=0.825\textwidth]{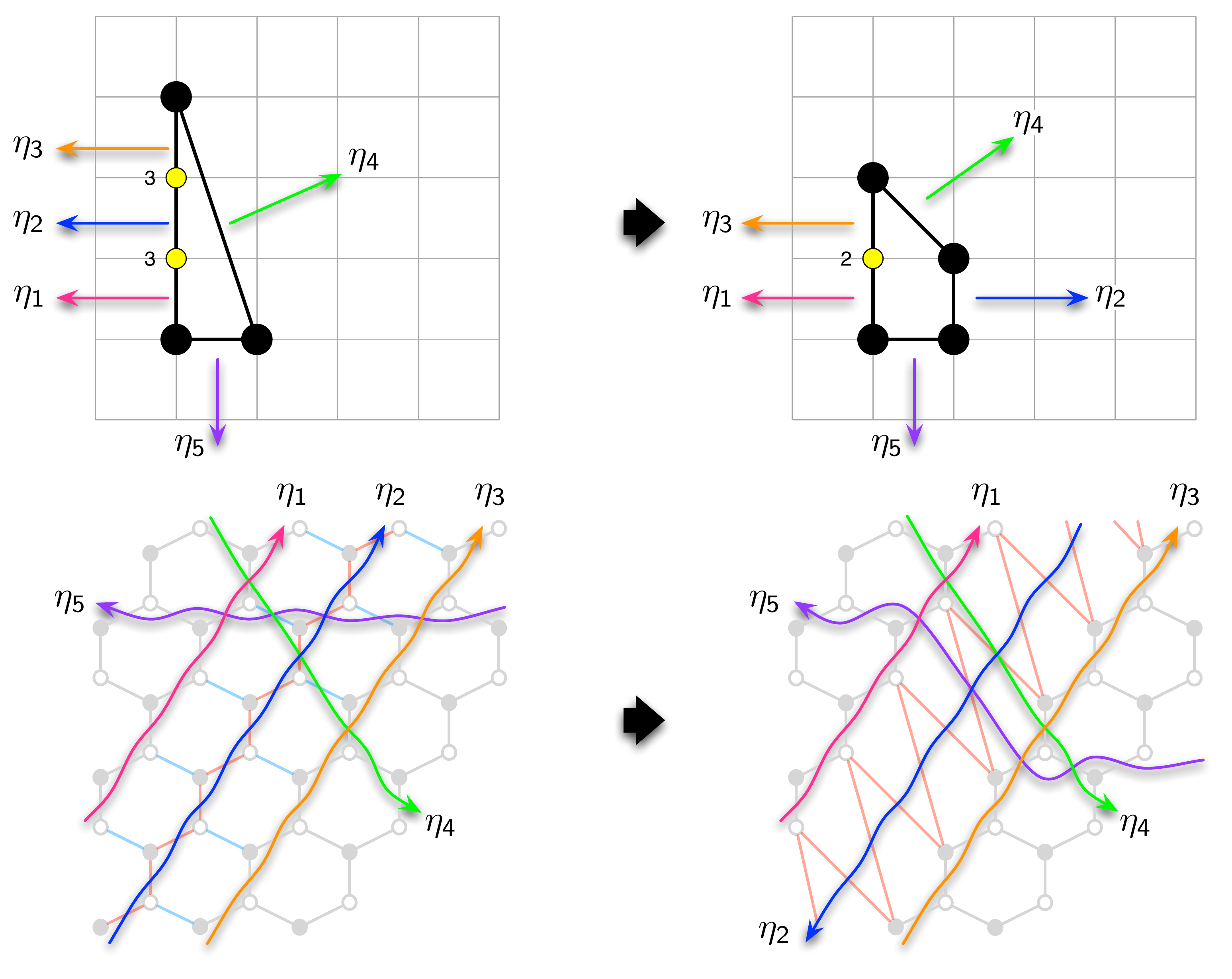} 
\caption{Mass deformation for the brane tiling of $\mathbb{C}^2/\mathbb{Z}_3 \times \mathbb{C}$ with zig-zag paths and the corresponding toric diagrams and external $(p,q)$-legs. The winding numbers of zig-zag paths in the brane tiling correspond to the directions of the $(p,q)$-legs. The toric diagram is the dual graph of $(p,q)$-web diagram and as such we can trace the effect of mass deformation on toric geometry.  \label{zigzagdef2}}
 \end{center}
 \end{figure}

It is instructive to follow other characteristics of brane tilings along a mass flow. \fref{zigzagdef} shows for the brane tiling of $\mathbb{C}^2/\mathbb{Z}_3\times \mathbb{C}$ how zig-zag paths transform from a generic UV brane tiling to a IR brane tiling. Zig-zag path $\eta_2$ reverses its direction due to the fact that all blue edges corresponding to fields given mass in the brane tiling get removed during the mass flow. The $(p,q)$-leg corresponding to the zig-zag path $\eta_2$ flips its direction, amounting to an area preserving deformation of the toric diagram as shown in \fref{zigzagdef2}. In fact, the new toric diagram is obtained from the old one by moving a single extremal toric point as illustrated in \fref{Z3case}. A few other zig-zag paths change their directions as well, but do not completely reverse their direction. This is governed by an overall $(p,q)$-charge conservation of zig-zag paths in the brane tiling.

In general, we observe for all examples presented in this work that mass flow of a brane tiling amounts to the displacement of toric points in the corresponding toric diagram and the reversal of $(p,q)$-leg directions in the corresponding dual $(p,q)$-web diagram. Furthermore, we remark that the area and hence the number of lattice points along the perimeter of the toric diagram of the UV and IR toric Calabi-Yau geometries are invariant because the mass deformation does not alter the number of anomalous and non-anomalous baryonic symmetries.\footnote{There are $e-3$ non-anomalous baryonic symmetries in the quiver, where $e$ is the number of lattice points along the perimeter of the toric diagram. There are also $2I$  anomalous baryonic symmetries, where $I$ is the number of points in the interior of the toric diagram. The number of gauge groups is $(e-3)+2I+1 = e+2I-2$, which by Pick's theorem is equal to twice the area of the toric diagram.}
\\

 \begin{figure}[t!]
\begin{center}
\includegraphics[trim=0cm 0cm 0cm 0cm,width=1\textwidth]{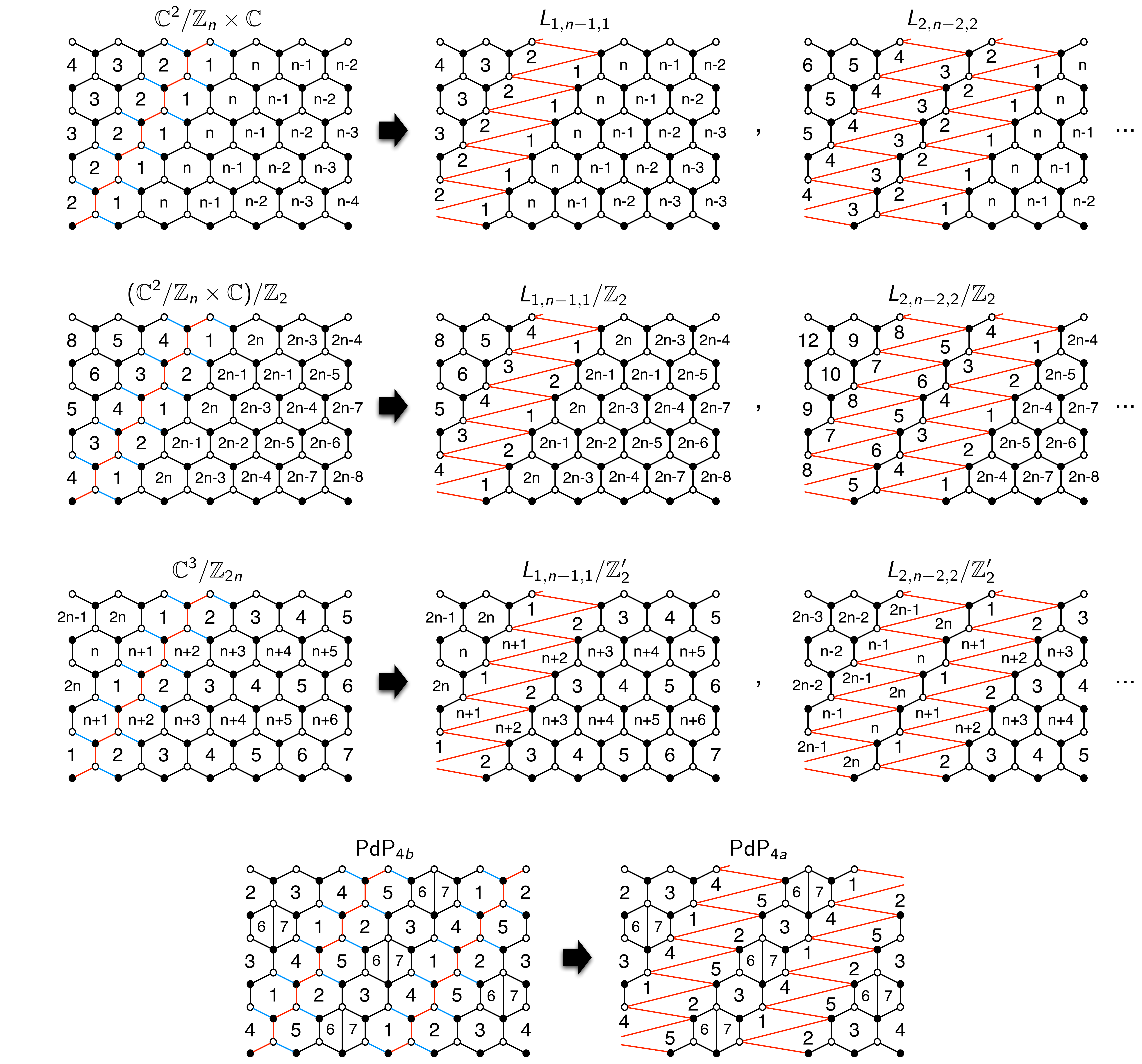}
\end{center}
\caption{Examples of mass flows with four/eight massive chiral multiplets.}
 \label{tflows}
\end{figure}

\subsection{Other flows}

\begin{figure}[H]
\begin{center}
\includegraphics[trim=0cm 0cm 0cm 0cm,width=0.9\textwidth]{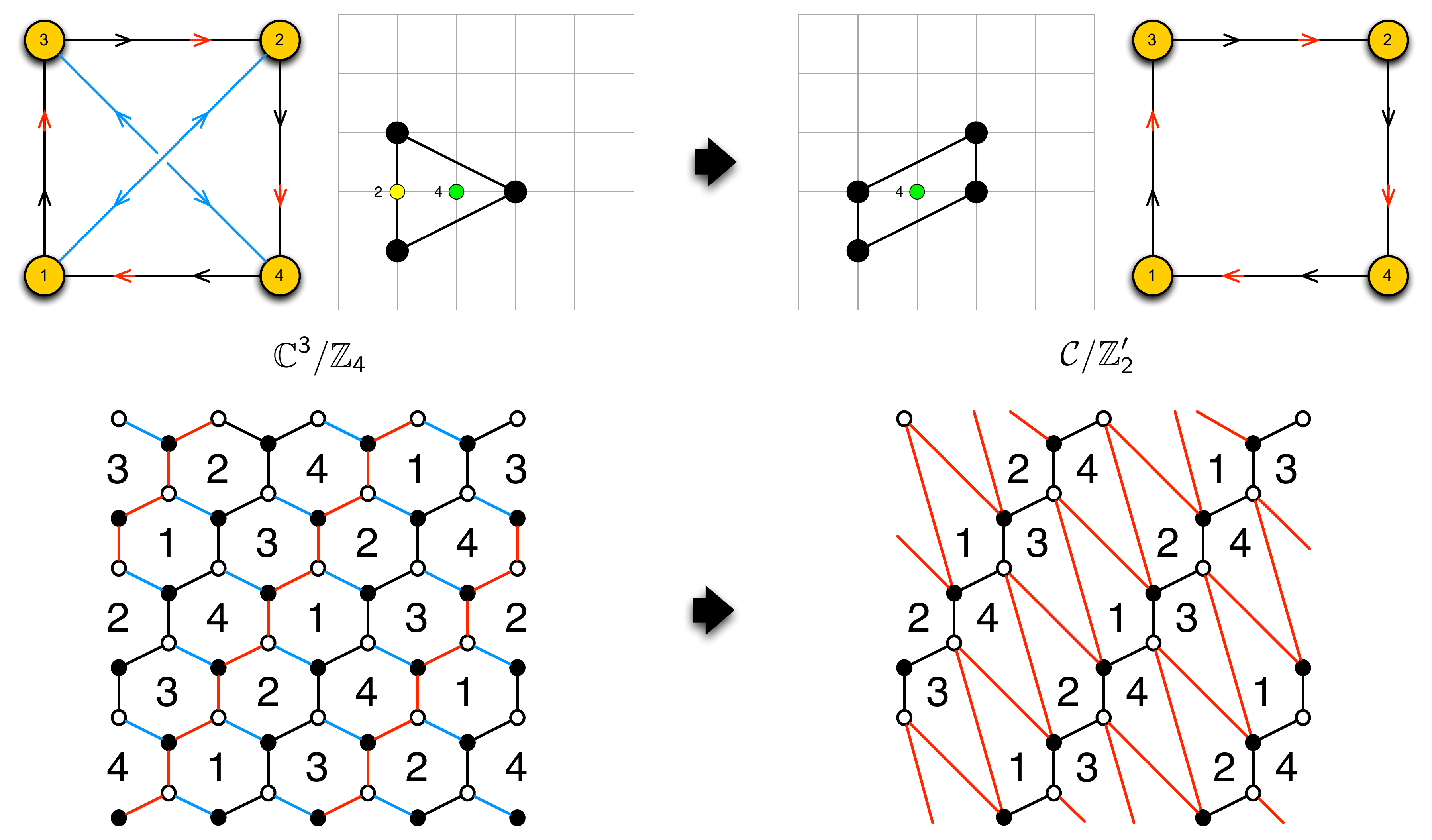}
\vspace{-0.7cm}
\end{center}
\caption{Mass flow from $\mathbb{C}^3/\mathbb{Z}_4$ to the second $\mathbb{Z}_2$ orbifold of the conifold $\mathcal{C}$, $\mathcal{C}/\mathbb{Z}_2^\prime$, which is equivalent to $L_{111}/\mathbb{Z}_2^\prime$.}
 \label{fflow43}
\end{figure}

\begin{figure}[H]
\begin{center}
\includegraphics[trim=0cm 0cm 0cm 0cm,width=0.9\textwidth]{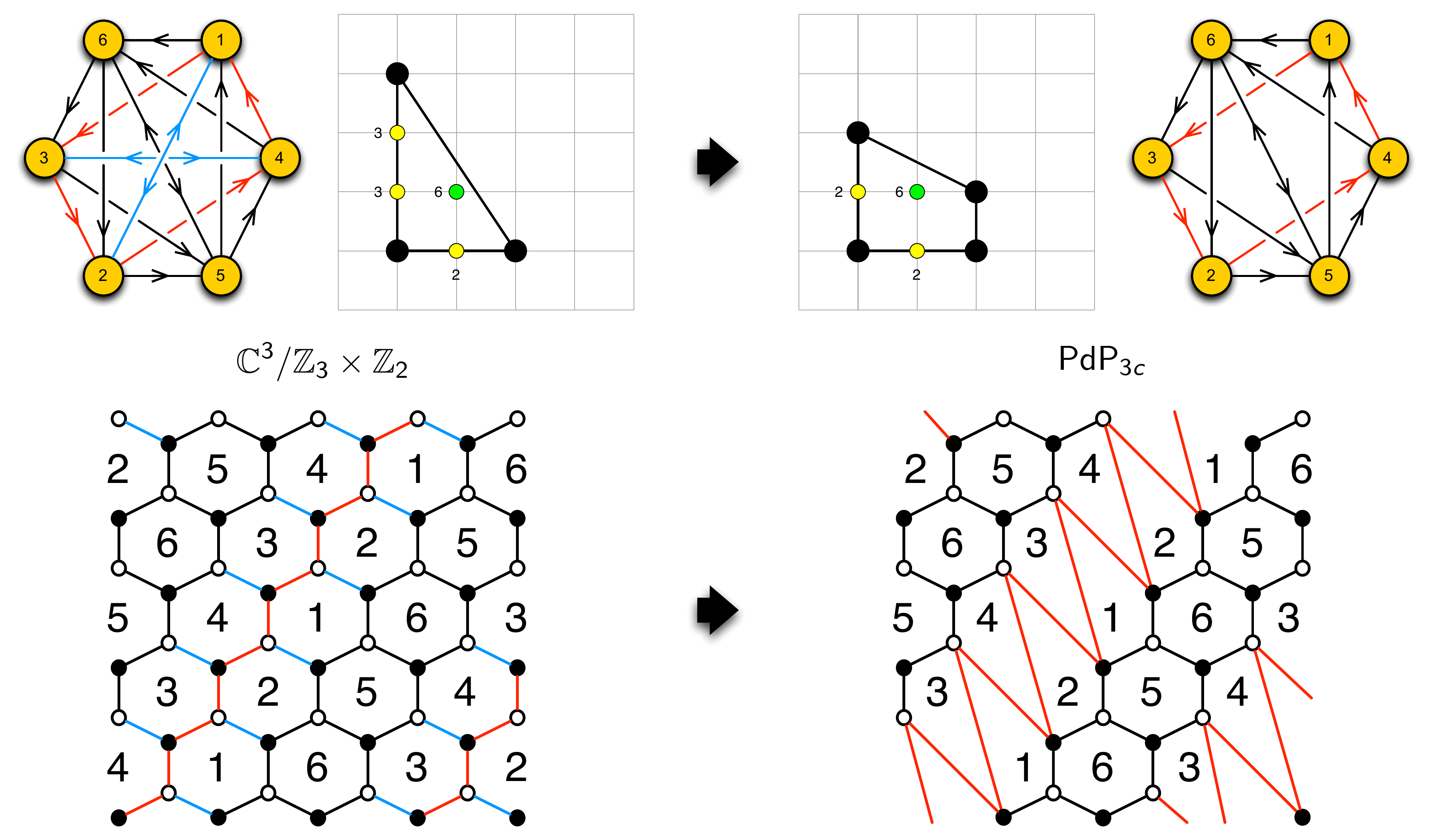}
\vspace{-0.7cm}
\end{center}
\caption{Mass flow from $(\mathbb{C}^2/\mathbb{Z}_3\times \mathbb{C})\times\mathbb{Z}_2$ to $L_{121}/\mathbb{Z}_2$ also known as $\text{PdP}_{3c}$.}
 \label{fflow61}
\end{figure}

\begin{figure}[H]
\begin{center}
\includegraphics[trim=0cm 0cm 0cm 0cm,width=0.9\textwidth]{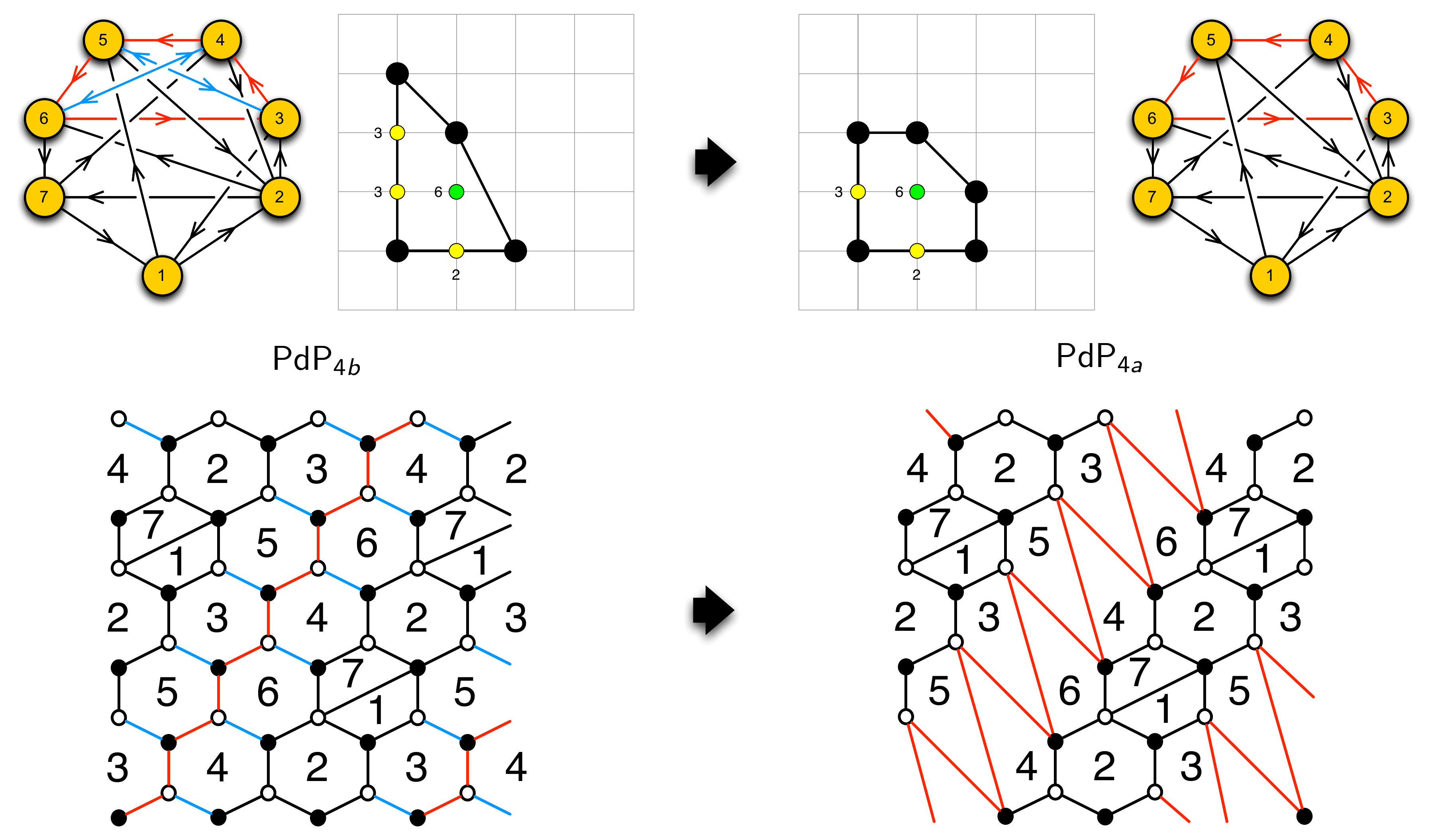}
\vspace{-0.7cm}
\end{center}
\caption{Mass flow from $\text{PdP}_{4b}$ to $\text{PdP}_{4a}$.}
 \label{fflow71}
\end{figure}

\begin{figure}[H]
\begin{center}
\includegraphics[trim=0cm 0cm 0cm 0cm,width=0.9\textwidth]{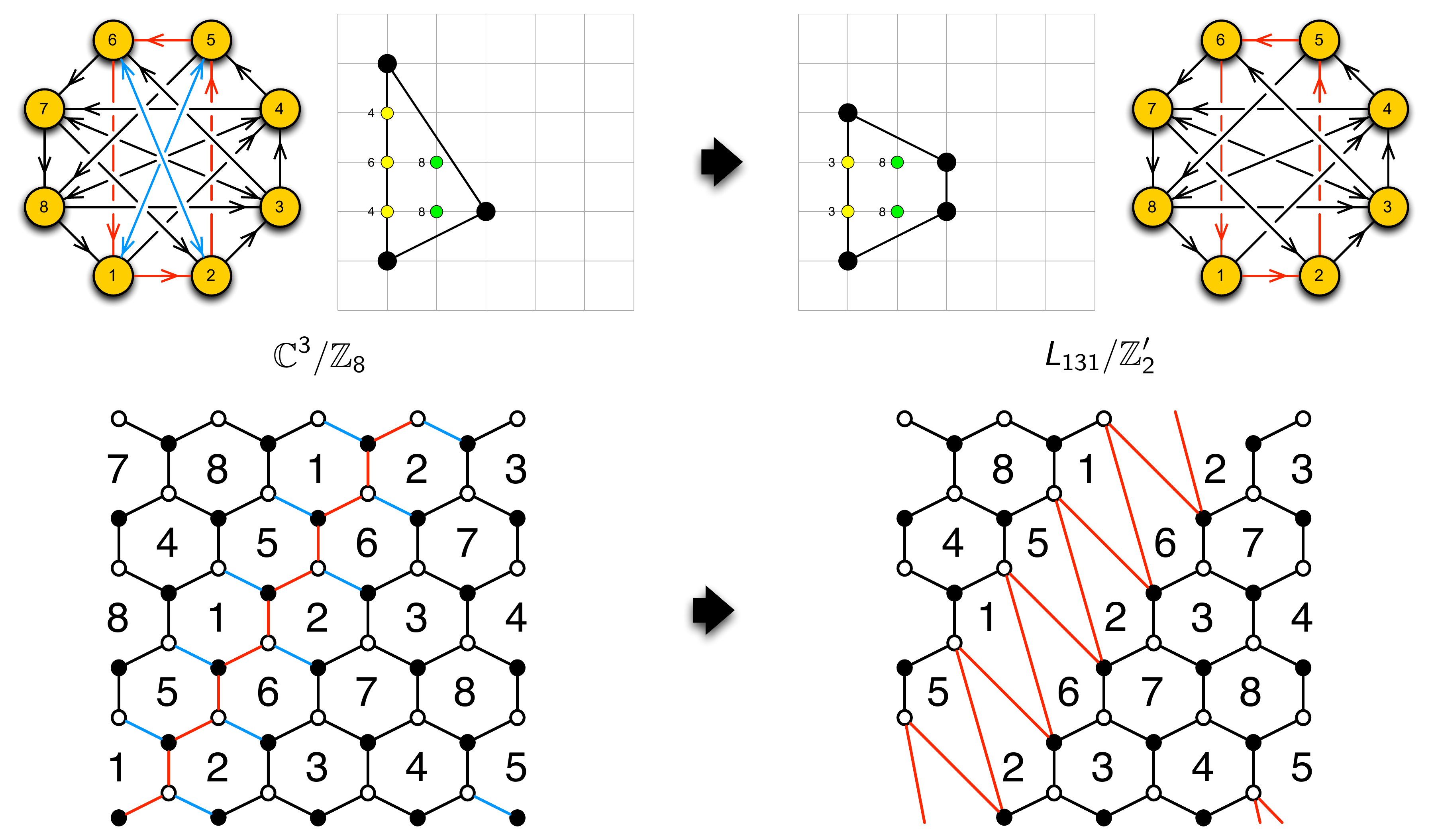}
\vspace{-0.7cm}
\end{center}
\caption{Mass flow from $\mathbb{C}^3/\mathbb{Z}_{8}$ to $L_{131}/\mathbb{Z}_2^\prime$.}
 \label{fflow80}
\end{figure}

\begin{figure}[H]
\begin{center}
\includegraphics[trim=0cm 0cm 0cm 0cm,width=0.9\textwidth]{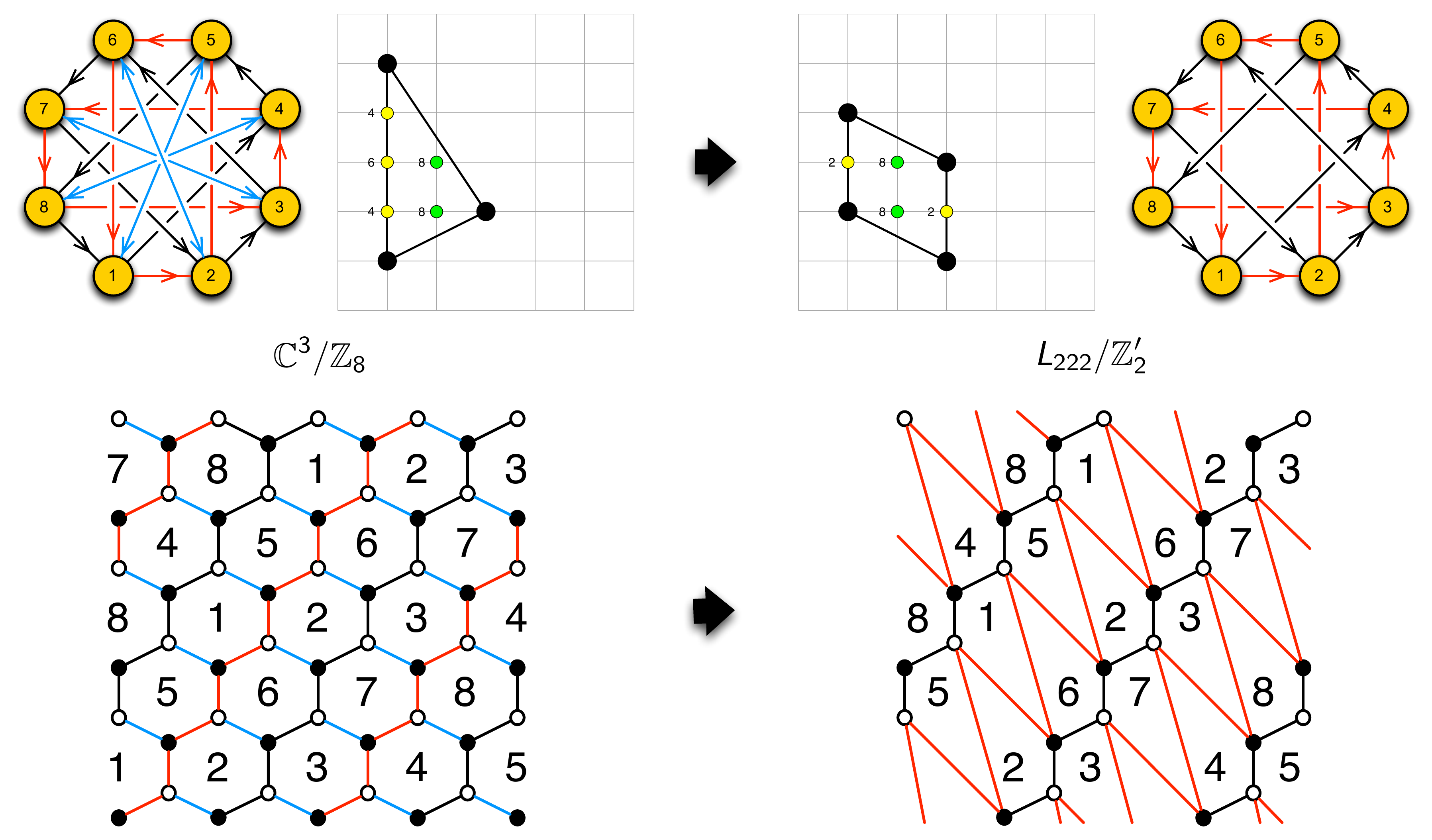}
\vspace{-0.7cm}
\end{center}
\caption{Mass flow from $\mathbb{C}^3/\mathbb{Z}_8$ to $L_{222}/\mathbb{Z}_2^\prime$.}
 \label{fflow80b}
\end{figure}

The analysis in the last section can be applied {\it mutatis mutandis} to a large class of brane tilings admitting mass deformations. In \fref{tflows}, we list three infinite sequences of flows starting from Abelian orbifolds of $\mathbb{C}^3$.  
\begin{itemize}
\item{ $\mathbb{C}^2/\mathbb{Z}_{n}\times \mathbb{C}  \to L^{k,n-k,k} $:  These flows are generated by giving identical masses to $k$ pairs of adjoint chiral multiplets which belong to $\mathcal{N} =2$ vector multiplets.  }
\item{ $(\mathbb{C}^2/\mathbb{Z}_{n}\times \mathbb{C})/\mathbb{Z}_{2} \to  L^{k,n-k,k}  /\mathbb{Z}_{2}  $:  These flows are generated by giving identical masses to $2k$ pairs of chiral multiplets in bifundamental representations.}
\item{ $\mathbb{C}^3/\mathbb{Z}_{2n}   \to L^{k,n-k,k} /\mathbb{Z}'_{2}  $:  These singularities admit mass deformations for $\mathbb{Z}_{2n}$ acting as $X^I\to \omega^{a_I} \, X^I$, with $\omega^{2n}=1$ and $(a_I) = (1,n-1,n)$ \cite{Davey:2010px,Hanany:2010ne,Hanany:2010cx,Davey:2011dd,Hanany:2011iw}. The flows are generated by giving identical masses to $2k$ pairs of  chiral multiplets in bifundamental representations.}
\end{itemize}
The results that we find on mass deformations of brane tilings can be extended to non-orbifold theories obtained via un-higgsing of orbifold theories \cite{Feng:2002fv,Hanany:2012hi}. As an example, we present in \fref{tflows} and \fref{fflow71} the flow starting from the brane tiling of $\mathrm{PdP}_{4b} $. The explicit form of the toric superpotentials, mass deformations, and the required field redefinitions for this example and all other examples are collected in appendices \sref{Lknk} to \sref{sapp4}. \fref{fflow41} to \fref{fflow80b} show in detail the brane tilings, toric diagrams and quiver diagrams for the first few examples of mass deformation that are classified in \fref{tflows}.
\\

 \begin{figure}[H]
\begin{center}
\includegraphics[trim=0cm 0cm 0cm 0cm,width=0.9\textwidth]{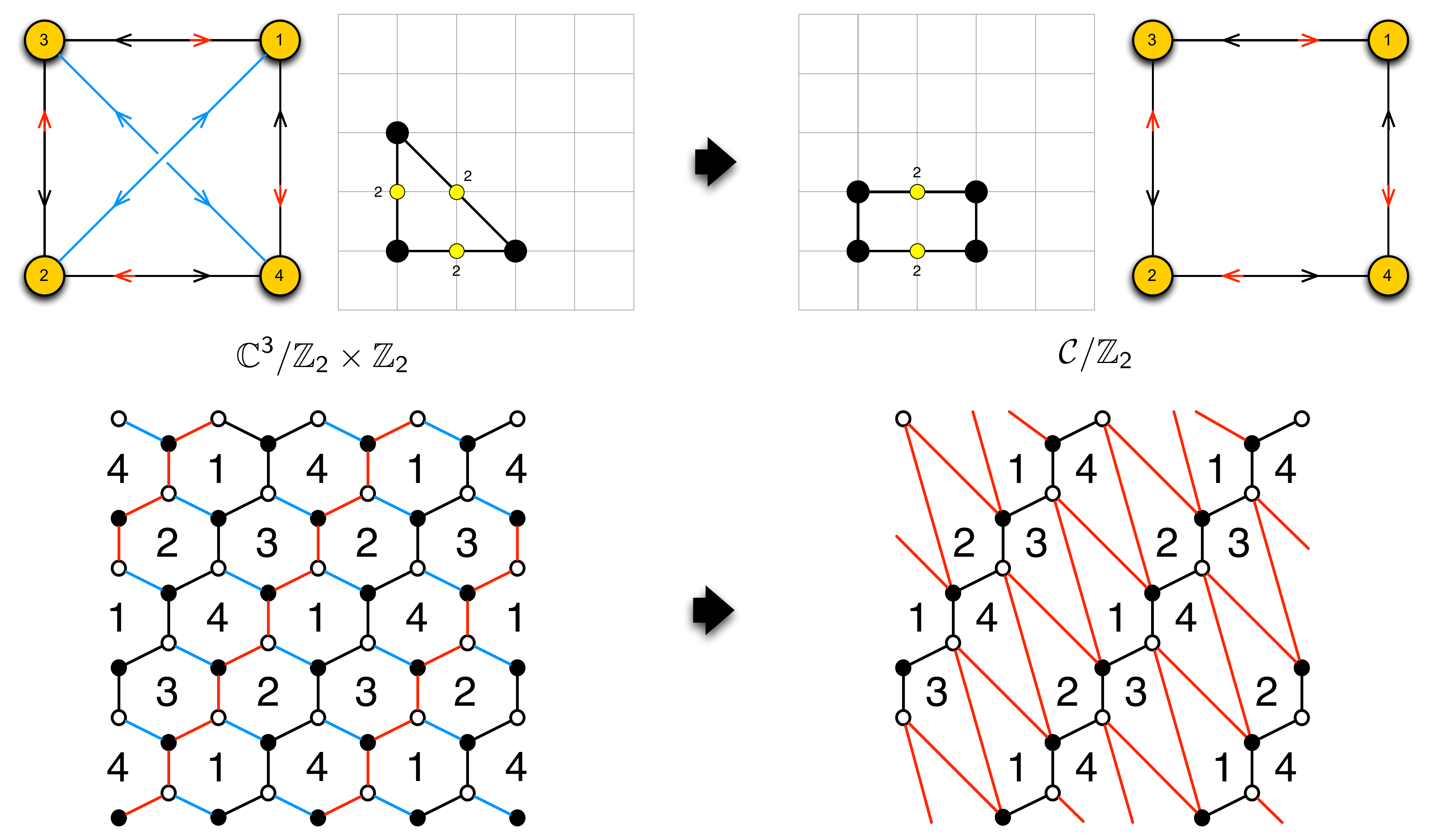}
\\
\vspace{-0.7cm}
\end{center}
\caption{Mass flow from $(\mathbb{C}^2/\mathbb{Z}_2\times \mathbb{C})\times\mathbb{Z}_2$ to $\mathcal{C}/\mathbb{Z}_2$ which is equivalent to $L_{111}/\mathbb{Z}_2$.}
 \label{fflow41}
\end{figure}

\section{R-symmetry, a-maximization and volume minimization \label{srch}}

It is well known that the R-symmetry in the $SU(2,2|1)$ superconformal algebra of any 4d SCFT is exactly and uniquely determined by an extremization procedure \cite{Intriligator:2003jj}. The exact superconformal R-symmetry maximizes the expression of the central charge $a$ \cite{Anselmi:1997am} in terms of 't Hooft anomalies
\be
a = {3\over 32} (3 Tr R^3 - Tr R) \label{aaa}~,
\ee
where the right-hand side is considered for any non-anomalous $U(1)$ R-symmetry $R_{trial}$ rather than just the superconformal one $R$.

In practice, one can start from a linear combination $R_{trial}=R_0+ \sum_I \alpha^i J_i$  of a fiducial non-anomalous R-symmetry generator $R_0$ with the non-anomalous Abelian global symmetry generators $J_i$, compute $a_{trial}$ and maximize with respect to $\alpha_i$. The maximum of $a_{trial}$ is  the central charge $a$ of the SCFT. As a corollary, one can argue that the central charge $a$ decreases along RG flows. Indeed, relevant deformations break some of the flavor symmetries so, in the absence of accidental symmetries, the extermination in the IR under a subset of the $\alpha_i$'s leads to a smaller  value $a_{IR}<a_{UV}$\cite{Intriligator:2003jj}.  For theories in this class, a generalization of \eref{aaa} has been introduced by \cite{Kutasov:2003ux} which can be used away from the endpoints of RG flows. The resulting $a$-function monotonically decreases  along the entire RG flow if there are no accidental symmetries. However, the RG flows under consideration in this paper exhibit accidental symmetries in the IR so we cannot use the a-function to study the RG flow locally in the energy scale.\footnote{One could use the $a$-function employed in the proof of the $a$-theorem by \cite{Komargodski:2011vj}, but that requires computing a scattering amplitude rather than a 't Hooft anomaly.} We will content ourselves with checking that $a_{IR}<a_{UV}$ in all RG flows under consideration, consistently with the $a$-theorem. 

We stress that, to obtain the correct superconformal R-charge and $a$ central charge of the infrared SCFT, it is essential to take into account the accidental mesonic symmetry that is made manifest by the field redefinitions discussed in the previous section. Maximizing only with respect to the mesonic symmetries that are present all along the RG flows generically leads to the wrong answer, as expected on general grounds \cite{Intriligator:2003jj}.

The connection with toric geometry and volume minimization was first pointed out in \cite{Martelli:2005tp}, where it is shown that the Reeb vector and the volume of a Sasaki-Einstein metric on the base of an $n$-dimensional toric Calabi-Yau cone may be computed by minimizing a function $Z$  which depends only on the toric data. For toric CY 3-folds, the Reeb vector and the volume correspond to the superconformal R-symmetry and the inverse of the central charge $c=a$ of the holographic dual SCFT, respectively. 
Agreement between volume minimization and a-maximization for toric SCFT's was shown in \cite{Butti:2005vn} and later generalized to non-toric cases in \cite{Butti:2006nk,Eager:2010yu}.
In the present investigation, we are interested in relevant mass deformations which, despite breaking the toric condition, lead to RG flows both of whose endpoints are toric SCFT's. We will show that for each of the considered flows, $V_{IR} > V_{UV}$ (or equivalently $a_{IR}< a_{UV}$ using $a$-maximization). 
\\

\subsection{Volume minimization}

We have already recalled that the mesonic moduli space is a Calabi-Yau cone $C(X)$ over a Sasaki-Einstein 5-manifold $X$ \cite{Benvenuti:2004dy,Franco:2005sm}. Let us compute first the volumes $V_{UV}$ and $V_{IR}$ of $X$ at the two ends of the mass flow. According to holography, the volumes $V$ and the Reeb vector can be found by extremizing a volume function $Z$  introduced in \cite{Martelli:2005tp}. The function $Z$ is encoded in the Hilbert series which counts chiral gauge invariant  operators of the SCFTs. More precisely, introducing  a fugacity $t_\alpha=e^{-\mu r_\alpha}$ for each of the GLSM field $p_\alpha$ 
associated to the CY singularity, the volume function $Z$ is defined as
   \beal{e90i01}
Z(r_\alpha; {\cal M} ) =  \lim_{\mu\rightarrow 0 } \mu^3 g (e^{-\mu r_\alpha}; {\cal M}  )~~.
\eea
Note that we are overparametrizing the space of R-charges: the volume function is invariant under 
\be\label{gauge}
r_\alpha \to r_\alpha + \sum_{i=1}^{c-3} s^i (Q_t)_{i\alpha}~,
\ee
since the mesonic moduli space is the K\"ahler quotient \eqref{e200n23}. The freedom \eqref{gauge} can be used to fix $c-3$ of the $c$ perfect matching variables. In addition, because each perfect matching variables appears exactly once in each superpotential term, which has R-charge $2$, the R-charges $r_\alpha$ satisfy 
\be\label{constraint_r_alpha}
\sum_{\alpha} r_\alpha=2~.
\ee
The remaining 2-dimensional subspace corresponds to the mixing of the R-symmetry with mesonic symmetries. Extremizing $Z$ with respect to $r_\alpha$ over this subspace leads to the volume
\be
V( {\cal M} ) = \Omega \cdot {\rm min}~  Z(r_\alpha; {\cal M} ) ~, \qquad \Omega  =\left(\frac{2\pi}{3}\right)^3~,
\ee
where we have introduced a suitable normalization factor $\Omega$.
The R-charges of the SCFT are obtained from the values of $ r_\alpha$ extremizing $Z$.  For example for ${\cal M}=\mathbb{C}^3$ 
 \be
  Z={1 \over r_1 r_2 (2-r_1 -r_2) }        \qquad \Rightarrow   \qquad V_{S^5} = \pi^3       \quad ~~~r_1=r_2={2\over 3} \label{zc3}
 \ee
as expected for the ${\cal N}=4$ theory. 
In the following paragraphs, we compute the volume of the SCFTs at the two ends of the mass flows under  consideration in this paper.
\\

\subsection*{Volumes for  $\mathbb{C}^2/\mathbb{Z}_n \times \mathbb{C} \to L^{k ,n-k, k}$  and their orbifolds}

We start by considering  the flow starting from $\mathbb{C}^2/\mathbb{Z}_n \times \mathbb{C}$.  
 The volume of an  orbifold of  a manifold ${\cal M}$ is simply the volume of $\cal M$ divided by the order of the group. 
For $S^5/\mathbb{Z}_n$ one then finds
 \be
  V_{S^5/\mathbb{Z}_n} ={\pi^3\over n} ~.    \label{omega5zn}     
\ee 

On the other hand the volume of $L^{a,b,a}$ can be obtained by extremizing the volume function\footnote{This expression follows easily from the toric description for $L^{a,b,a}$: the toric diagram has vertices $w_1=(0,0)$, $w_2=(1,0)$, $w_3=(1,a)$, $w_4=(0,b)$. The charge matrix of the GLSM is $Q_t=(-a,b,-b,a)$. The singularity is the hypersurface $xy=z^a w^b$ in $\mathbb{C}^4$. The computation here agrees with the results in \cite{Benvenuti:2004dy,Franco:2005sm}.} 
 \be
 Z_{L^{a,b,a}}=  {b r_1+ar_2+ar_3+b r_4 \over (b r_1+a r_2)(a r_3+b r_4)(r_2+r_3)(r_1+r_4)}  ~,
 \ee
with 
\be
\begin{split}
&r_1+r_2+r_3+r_4=2 \\ (r_1,r_2,r_3,r_4) &\sim (r_1,r_2,r_3,r_4) + (-a,b,-b,a)s~.
\end{split} 
\ee
Extremizing along the 2-dimensional subspace corresponding to mixing with the mesonic symmetries, one finds
 \be
  V_{L^{a,b,a}}= \frac{4\pi^3}{27 a^2 b^2}  \left[(2b-a)(2a-b)(a+b) +2 (a^2+b^2-ab)^{3\over 2} \right] 
 \ee
Taking $a=k$ and $b=n-k$, one can check that  $V_{L^{k,n-k,k}}> V_{S^5/\mathbb{Z}_n}  $ for all $n>0$ and $k=1,\dots,\left[\frac{n}{2}\right]$ (which covers the entire range because $L^{a,b,a}=L^{b,a,b}$). As expected, the volume of the Sasaki-Einstein manifold increases along the RG flow. Similarly, for the volumes of $  \mathbb{C}^3/\mathbb{Z}_n \times \mathbb{Z}_2$ and  $ \mathbb{C}^3/\mathbb{Z}_{2n}  $ 
 one finds half the result \eref{omega5zn} since the order is twice as larger.  On the other hand, the endpoints of the flows starting on these singularities are $\mathbb{Z}_2$ orbifolds of $L^{k,n-k,k}$ and therefore the volumes are just half of  those of $L^{k,n-k,k}$.  Again the volumes increase along the flow as expected.
\\

 \subsection*{Volumes flow for $\mathrm{PdP}_{4b} \to \mathrm{PdP}_{4a}$}

Finally, for $\mathrm{PdP}_{4b}$ one finds\footnote{The toric diagram of  $\mathrm{PdP}_{4b}$ has vertices $w_1=(0,0)$, $w_2=(1,0)$, $w_3=(2,1)$, $w_4=(0,3)$. The charge matrix of the associated GLSM is $Q_t=(-4,6,-3,1)$. $\mathrm{PdP}_{4b}$ is a non-complete intersection. The computation here agrees with the results in \cite{Hanany:2012hi}.}
\beal{es6}
Z_{\mathrm{PdP}_{4b}} = \frac{3 r_1 + 2 r_2 + 4 r_3 + 12 r_4}{(r_3 + 3 r_4) (4 r_4 + r_1) (2 r_3 + r_2) (3 r_1 + 2 r_2)}
~~,
\eea
with 
\be
\begin{split}
&r_1+r_2+r_3+r_4=2 \\ (r_1,r_2,r_3,r_4) &\sim (r_1,r_2,r_3,r_4) + (-4,6,-3,1)s~.
\end{split} 
\ee
 The volume is given by the minimum
\beal{es6}
V_{\mathrm{PdP}_{4b}} \simeq 0.531049 \, \Omega ~~.
\eea
In comparison, $\mathrm{PdP}_{4a}$  has the volume function\footnote{The toric diagram of $\mathrm{PdP}_{4a}$ has vertices $w_1=(0,2)$, $w_2=(1,2)$, $w_3=(0,0)$, $w_4=(2,1)$ and $w_5=(2,0)$. The charge matrix of the associated GLSM is $Q=\begin{pmatrix} 2 & -2 & -1 & 0 & 1\\ 1 & 0 & -1 & -2 & 2\end{pmatrix}$. The computation here agrees with the results in \cite{Hanany:2012hi}.}
\beal{es7}
&&
Z_{\mathrm{PdP}_{4a}} = 
(4 r_1^2 + 2 r_2^2 + 6 r_1 r_2 + 16 r_1 r_3 + 8 r_1 r_4 + 12 r_1 r_5 +  12 r_2 r_3  + 5  r_2 r_4 + 8 r_2 r_5 
\nn\\
&&
\hspace{0.5cm}
+ 12 r_3^2 + 12 r_3 r_4 +2 r_4^2 + 16 r_3 r_5 + 6 r_4 r_5 + 4 r_5^2)
\times
\nn\\
&&
\hspace{0.5cm}
\frac{1
}{
(2 r_1 + r_2 + 2 r_3) (2 r_1 + 2 r_2 + r_4) (r_1 + 3 r_3 + r_5)  (r_2 + 2 r_4 + 2 r_5) (2 r_3 + r_4 +  2 r_5)
}
~~,
\nn
\\
\eea
with 
\be
\begin{split}
&r_1+r_2+r_3+r_4+r_5=2 \\ (r_1,r_2,r_3,r_4,r_5) &\sim (r_1,r_2,r_3,r_4,r_5) + (2,-2,-1,0,1)s_1+(1,0,-1,-2,2)s_2~,
\end{split} 
\ee
leading to the volume
\beal{es8}
V_{\mathrm{PdP}_{4a}} \simeq
0.595008\,  \Omega~.
\eea
The volume of the Sasaki-Einstein manifold increases in agreement with the holographic $a$-theorem.
\\

\subsection{Volume ratios}

\begin{figure}[ht!!]
\begin{center}
\includegraphics[trim=0cm 0cm 0cm 0cm,width=0.3\textwidth]{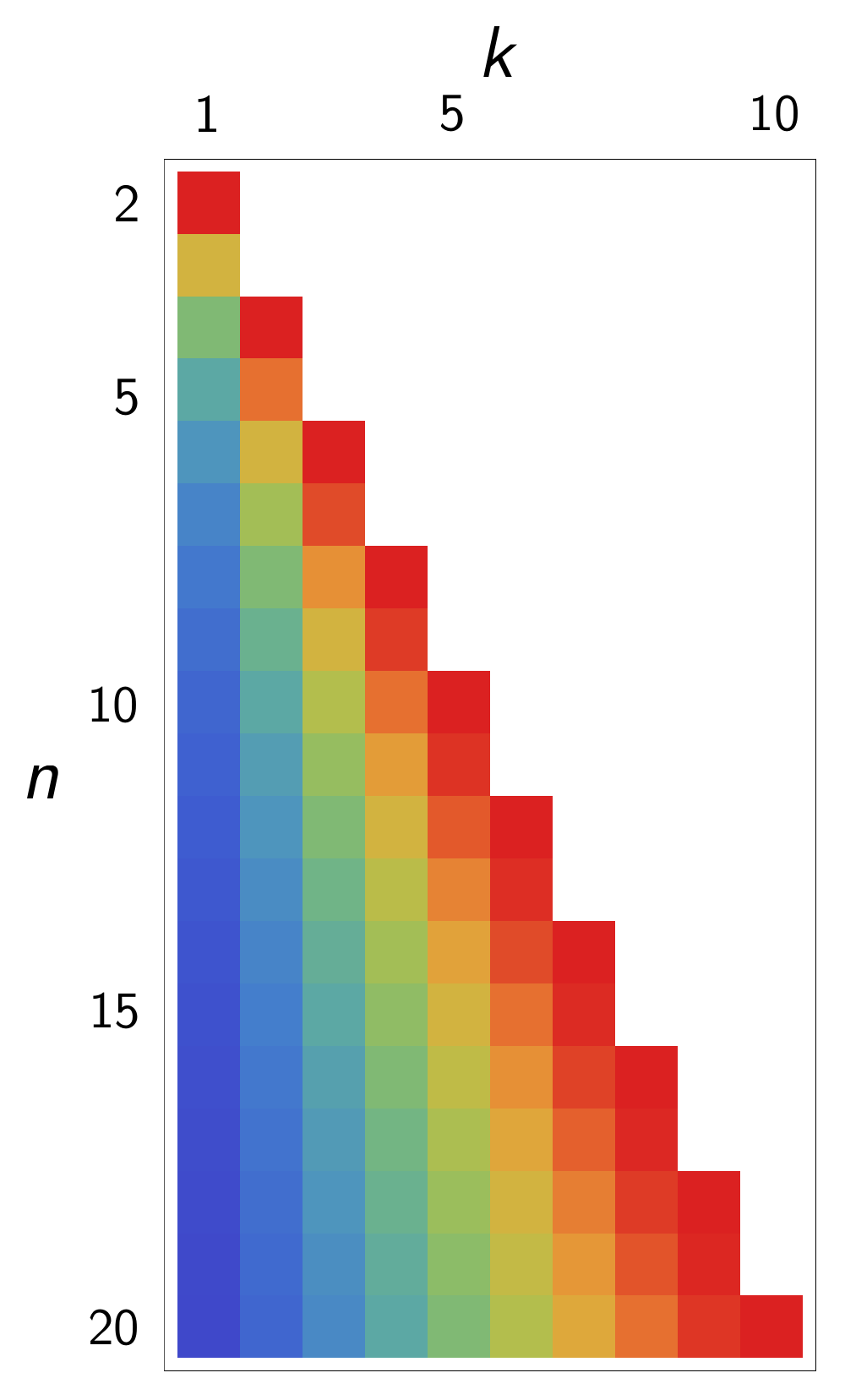}
\includegraphics[trim=0cm 0cm 0cm 0cm,width=0.68\textwidth]{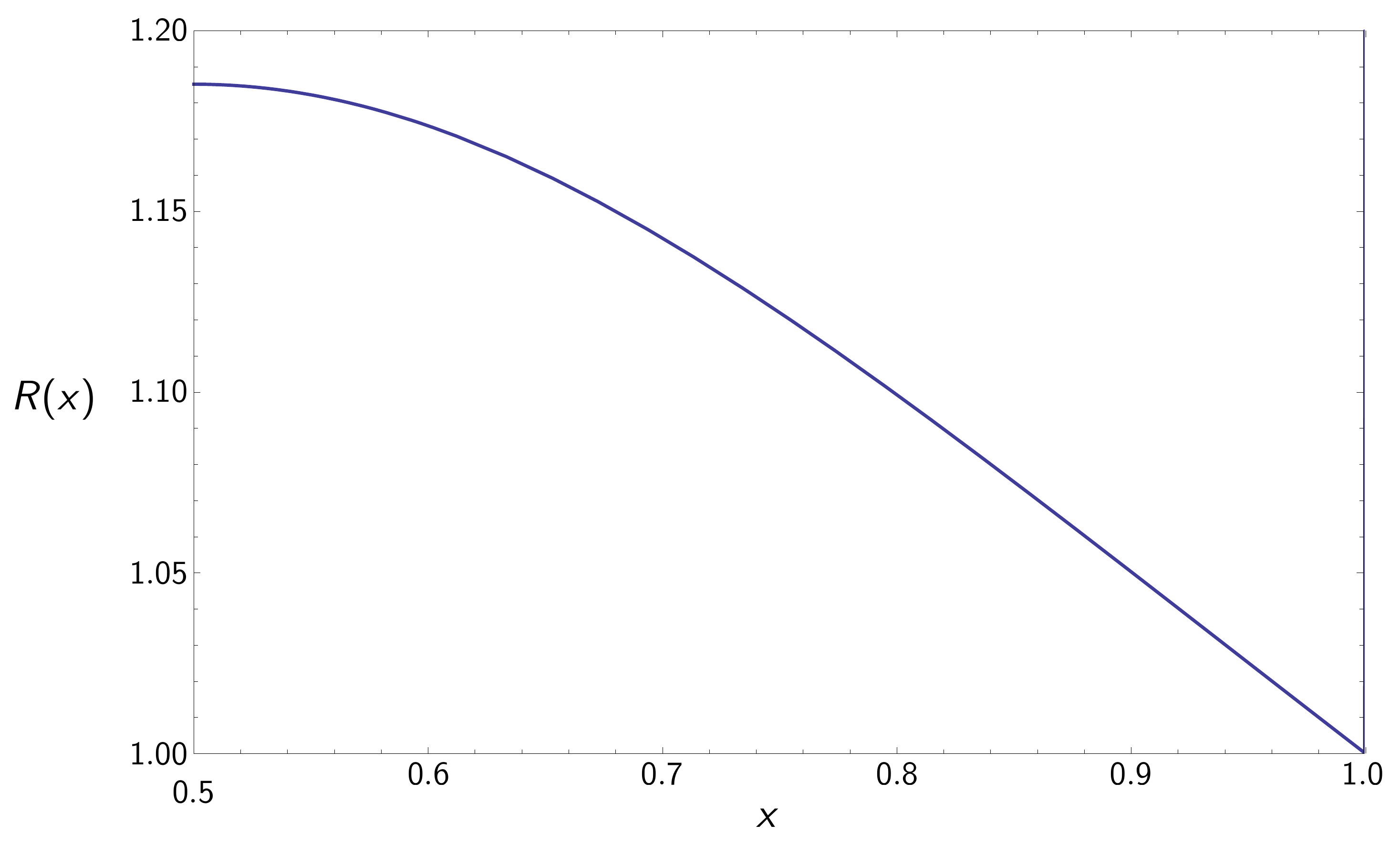}
\caption{
Volume ratio plot for mass flows $\mathbb{C}^2/\mathbb{Z}_{n}\times \mathbb{C}  \to L^{k,n-k,k}$, $(\mathbb{C}^2/\mathbb{Z}_{n}\times \mathbb{C})/\mathbb{Z}_{2} \to  L^{k,n-k,k}  /\mathbb{Z}_{2}  $ and $\mathbb{C}^3/\mathbb{Z}_{2n}   \to L^{k,n-k,k} /\mathbb{Z}'_{2}$.
Different colors correspond to different values of the volume ratio $R(k/n)$ in the left plot. The right plot shows the correspondence between the volume ratio $R(x)$ and the flow parameter $x=\frac{k}{n}$.
\label{fvol}
}
 \end{center}
 \end{figure}

Interesting observations can be made, when we consider ratios of volumes of the Sasaki-Einstein manifolds along the mass flow. The volume ratio of the mass deformations in \fref{tflows} are as follows,
\beal{esvol1}
\frac{V_{L^{k,n-k,k}}}{V_{\mathbb{C}^2/\mathbb{Z}_n\times \mathbb{C}}}
&=&
\frac{V_{L^{k,n-k,k}/\mathbb{Z}_2}}{V_{(\mathbb{C}^2/\mathbb{Z}_n\times \mathbb{C})/\mathbb{Z}_2}}
=
\frac{V_{L^{k,n-k,k}/\mathbb{Z}_2^\prime}}{V_{\mathbb{C}^2/\mathbb{Z}_{2n}}}
= R(k/n) 
~,~
\eea
where 
\beal{evrat}
R(x)
=
\frac{
4 (-9 x^2  + 9 x - 2 + 2 (3 x^2 - 3 x + 1)^{3/2})
}{
27 x^2 (1 - x)^2
}
~.~
\eea
We note that the volume ratios $R(x)$ in \eref{evrat} for the mass flows $\mathbb{C}^2/\mathbb{Z}_{n}\times \mathbb{C}  \to L^{k,n-k,k}$, $(\mathbb{C}^2/\mathbb{Z}_{n}\times \mathbb{C})/\mathbb{Z}_{2} \to  L^{k,n-k,k}  /\mathbb{Z}_{2}  $ and $\mathbb{C}^3/\mathbb{Z}_{2n}   \to L^{k,n-k,k} /\mathbb{Z}'_{2}  $ take particular irrational values for various values of $x=\frac{k}{n}$. The first few values are given in \tref{tvolrat} and plotted in \fref{fvol}.

\begin{table}[ht!]
\begin{center}
\begin{tabular}{|ll|cccc|}
\hline
& & \multicolumn{4}{c|}{$k$}
\\
&  & 1 & 2 & 3 & 4
\\
\hline
\multirow{8}{*}{$n$} & 2 &  $\frac{32}{27}$ & & & 
\\
& 3 & $\frac{2}{\sqrt{3}}$ & & & 
\\
& 4 & $\frac{32}{243} \left(7 \sqrt{7} - 10 \right)$ & $\frac{32}{27}$  & &
\\
& 5 & $\frac{5}{54} \left(13 \sqrt{13}-35 \right)$ & $\frac{10}{243} \left(7 \sqrt{7}+ 10\right)$ & & 
\\
& 6 & $\frac{16}{75} \left(7 \sqrt{21} -27 \right)$ & $\frac{2}{\sqrt{3}}$ & $\frac{32}{27}$ & 
\\
& 7 & 
$\frac{7}{243} \left(62 \sqrt{31}-308\right)$ &
$\frac{7}{675} \left(38 \sqrt{19}-56\right)$ &
$\frac{7}{486} \left(35+13 \sqrt{13}\right)$ & 
\\
& 8 &
$\frac{64}{1323} \left(43 \sqrt{43}-260\right)$ &
$\frac{32}{243} \left(7 \sqrt{7}-10\right)$ &
$\frac{64}{6075} \left(28+19 \sqrt{19}\right)$ &
$\frac{32}{27}$
\\
& 9 &
$\frac{1}{8} \left(19 \sqrt{57}-135\right)$ &
$\frac{2}{49} \left(13 \sqrt{39}-54\right)$ &
$\frac{2}{\sqrt{3}}$ &
$\frac{1}{50} \left(27+7 \sqrt{21}\right)$ 
\\
\hline
\end{tabular}
\caption{
Values of volume ratios $R(k/n)$ for various values of $n$ and $k=1,\dots,[n/2]$ for mass flows $\mathbb{C}^2/\mathbb{Z}_{n}\times \mathbb{C}  \to L^{k,n-k,k}$, $(\mathbb{C}^2/\mathbb{Z}_{n}\times \mathbb{C})/\mathbb{Z}_{2} \to  L^{k,n-k,k}  /\mathbb{Z}_{2}  $ and $\mathbb{C}^3/\mathbb{Z}_{2n}   \to L^{k,n-k,k} /\mathbb{Z}'_{2}  $.
\label{tvolrat}
}
\end{center}
\end{table}

Since the volume ratio $R$ only depends on $x=\frac{k}{n}$, rescaling $n$ and $k$ by integers provides an infinite class of mass flows which are characterized by the same ratio $R(x)$. This phenomenon has been observed for the mass flow of $\mathbb{C}^2/\mathbb{Z}_{n}\times \mathbb{C}$ in the very special case of $n=2k$, i.e. $x=\frac{1}{2}$. This corresponds to the case where one gives masses to all adjoints fields in the theory and one finds the universal value $32/27$ for the ratio (see \tref{tvolrat}). 
This is consistent with the findings in \cite{Tachikawa:2009tt}, where it was shown that when a $\mathcal{N}=2$ SCFT flows to a $\mathcal{N}=1$ SCFT by masses given to all adjoint fields, the central charges $a$ and $c$ before and after the flow take the ratio $R=\frac{32}{27}$. 

Interestingly, the ratio $R=\frac{32}{27}$ is the maximum value achieved by the flows considered in this paper. Here we have extended this observation beyond masses given to adjoints and with the result that there is an infinite number of `mass flow classes'  characterized by their ratios $R(x)$ with $x\in[\frac{1}{2},1)$. This is shown for mass flows $\mathbb{C}^2/\mathbb{Z}_{n}\times \mathbb{C}  \to L^{k,n-k,k}$, $(\mathbb{C}^2/\mathbb{Z}_{n}\times \mathbb{C})/\mathbb{Z}_{2} \to  L^{k,n-k,k}  /\mathbb{Z}_{2}  $ and $\mathbb{C}^3/\mathbb{Z}_{2n}   \to L^{k,n-k,k} /\mathbb{Z}'_{2}$. It would be interesting to study this phenomenon further in future work.

\section{Mass deformations as complex structure deformations \\ and Hilbert series \label{sflow}}

The introduction of masses deforms the superpotential of  the toric SCFT and therefore its F-term relations. In this section, we show how this results into a complex structure deformation of the underlying singularity by looking directly at the algebraic description of the mesonic moduli space and its Hilbert series,
rather than performing field redefinitions of the microscopic fields. 

Let us consider the simplest flow 
 $\mathbb{C}^2/\mathbb{Z}_n\times \mathbb{C}  \to L^{k, n-k,k}$ as an illustration of the general phenomenon. The gauge invariant mesonic operators are given by  
  \be
     x=\prod_{i=1}^n  X_{i,i+1}  \quad , \quad   y=\prod_{i=1}^n  X_{i,i+1}   \quad , \quad  w_i=X_{i,i+1} X_{i+1, i}  \quad , \quad \phi_i    \quad \quad
  \ee
  satisfying
  \be
   x \,y =\prod_{i=1}^n w_i  ~.  \label{reli}
  \ee
After the mass deformation the relations among these operators that follow from their definition and the F-term equations become\footnote{We restrict to the mesonic branch of the moduli space, where the adjoint fields are equal. This condition is not satisfied on the branches of the moduli space corresponding to the regular D3-brane splitting into fractional D3-branes.}
   \bea
   && \phi_i=z   ~, \qquad \forall i=1,...,n \nn\\
   && w_{2i-2}-w_{2i-1}+m \, \phi_{2i-1}= w_{2i-1}-w_{2i}-m \, \phi_{2i}=0~,    \qquad i=1,...,k \nn\\
   && w_{i-1}-w_{i}+m \, \phi_i=0   ~, \qquad i=2k+1,...,n~. 
   \eea
  The solution can be written as
  \bea
   w_{2i-1}-m z=w_{2i}=w   \qquad \qquad i=1,...,k \nn\\
   w_i=w  \qquad \qquad i=2k+1,...,n 
   \eea
     while the relation \eref{reli} becomes 
   \be\label{def_alg}
    x \, y=(w+m\, z)^k \, w^{n-k}~.
   \ee  
If $m=0$, \eqref{def_alg} reduces to the algebraic description of $\mathbb{C}^2/\mathbb{Z}_{n}$ in terms of the generators $x$, $y$ and $w$, which is in a product with the $\mathbb{C}$ plane generated by $z$. If $m \neq 0$, the equation which describes the singularity is deformed, and the symmetry acting on $x$, $y$, $z$ and $w$ by complex rescalings is broken from $\mathbb{C}^{*3}$ to $\mathbb{C}^{*2}$. However, we can change variable $z$ to $\tilde{z}=mz+ w$ in order to recast \eqref{def_alg} into
\be\label{def_alg_2}
    x \, y = \tilde{z}^k \, w^{n-k}~.
   \ee
This is nothing but the defining equation of the cone over $L^{k,n-k,n}$, which enjoys the full toric $\mathbb{C}^{*3}$ symmetry. We see that as soon as the mass parameter is turned on, the complex structure of the mesonic moduli space becomes that of the cone over $L^{k,n-k,n}$, which is associated to the infrared SCFT. This fact is well known for the flow  $\mathbb{C}^2/\mathbb{Z}_n\times \mathbb{C}  \to L^{k, n-k,k}$ \cite{Gubser:1998ia}, although here we have emphasized the choices of mass parameters that lead to a toric mesonic moduli space for the mass-deformed theory (or equivalently in the infrared). 
We claim that this phenomenon, which was studied from the dual supergravity perspective in \cite{Halmagyi:2004jy}, applies to all the mass flows described in this paper even in the cases where the singularities are not complete intersections.  

There are two ways to understand why the mesonic moduli space becomes the toric Calabi-Yau threefold associated to the infrared field theory as soon as the mass deformation is turned on. The first way is to analyze directly the $F$-term equations, as reviewed in section \sref{s2} and detailed in Appendix A. Since imposing $F$-term equations effectively integrates out massive fields, the mesonic moduli spaces of the mass-deformed UV theory and of the low energy theory obtained by integrating out the massive fields coincide. 
Using the light matter fields of the UV theory only manifests a $U(1)^2$ symmetry in the mesonic moduli space, because the mass deformation explicitly breaks a $U(1)$ factor of the toric $U(1)^3$ of the UV CFT. However a change of variables \eqref{e200n40} recasts the low energy superpotential in toric form, making it clear that the mesonic moduli space of the IR CFT (or, equivalently, of the mass-deformed UV CFT) is a toric Calabi-Yau threefold.

An alternative and more general perspective on this point, even though it misses the accidental symmetry, is offered by the Hilbert series of the mesonic moduli space of the Abelian quiver gauge theory \cite{Benvenuti:2006qr,Feng:2007ur}. In this section we work with the matter fields of the quiver gauge theory subject to $F$-term equations, rather than with the perfect matching variables appearing in \eqref{e200n30}, that are associated to the GLSM description of the toric Calabi-Yau, which has a larger gauge group and no superpotential. 
The Hilbert series of the mesonic moduli space counts gauge invariant chiral operators, weighted according to their charges under the global symmetry. 
It is given by a Molien formula similar to \eqref{e200n30}, but now the integral is over the gauge group of the quiver gauge theory. Chiral multiplets $\Phi$ in the quiver contribute to the integrand factors $(1-t_\Phi)^{-1}$, where $t_\Phi$ is the weight of $\Phi$ under the global and gauge symmetry group, whereas $F$-term equations contribute to the numerator. 
To proceed, we note that the $F$-term equations for the massive fields are linear in the massive fields and can be solved independently of the other equations. So the $F$-term of a massive field $X$ contributes a factor $(1-t_{\frac{\partial W}{\partial X}})$ to the numerator of the integrand.

We start from the Hilbert series of the mesonic moduli space of the toric UV theory, which depends on three fugacities associated to the $U(1)^3$ non-baryonic symmetry. All the weights above are monomials in such fugacities and the fugacities for the gauge group. 
The first effect of the mass deformation on the Hilbert series of the mesonic moduli space of the UV theory is to unrefine it with respect to the non-baryonic symmetry which is broken by the mass terms: the fugacity associated to the broken $U(1)$ symmetry is set to $1$.
Secondly, in the Molien integrand the $F$-term of a field $X$ appearing in a superpotential mass term $m X Y$ exactly cancels the contribution of the partner field $Y$, 
\be\label{cancellation}
\frac{1-t_{\frac{\partial W}{\partial X}}}{1-t_Y} = 1~,
\ee
because $\frac{\partial W}{\partial X}$ and $Y$ are forced by the mass term to have the same quantum numbers under the unbroken symmetries. 

Since massive fields and their $F$-term equations cancel out in the Hilbert series, we conclude that the Hilbert series of the mass-deformed UV theory coincides with the Hilbert series of the IR theory, unrefined with respect to the accidental symmetry. In particular, a necessary condition for two toric Calabi-Yau threefolds (complete intersections or not) to be related by a mass flow is that their Hilbert series coincide under a certain unrefinement. This is a restrictive constraint, because it implies a bijection between spectra of holomorphic functions.

As an example, let us return to the flows $\mathbb{C}^2/\mathbb{Z}_n\times \mathbb{C}  \to L^{k, n-k,k}$. 
For $\mathbb{C}^2/\mathbb{Z}_n\times \mathbb{C}$, parametrized by $(x,y,w,z)$ subject to $x y = w^n$, the refined Hilbert series is
\be
g(t; \mathbb{C}^2/\mathbb{Z}_n\times \mathbb{C}) = \PE \left[ t_x + t_y + t_w + t_z - t_w^n\right]~, \qquad \mathrm{with} \quad  t_x t_y = t_w^n~. 
\ee
Unrefining with respect to the broken $U(1)$ symmetry sets $t_z = t_w$ and yields 
\be\label{unref_orbifold}
g(t; \mathbb{C}^2/\mathbb{Z}_n\times \mathbb{C})|_{t_z = t_w}=\PE [ t_x + t_y + 2 t_w - t_w^n]~, \qquad \mathrm{with} \quad t_x t_y = t_w^n~.
\ee
For the cone over $L^{k, n-k,k}$, parametrized by $(X,Y,W,Z)$ subject to $X Y = Z^k W^{n-k}$, the refined Hilbert series is 
\be
g(T; L^{k, n-k,k}) = \PE \left[ t_X + t_Y + t_W + t_Z - t_Z^k t_W^{n-k}\right]~, \qquad \mathrm{with} \quad t_X t_Y = t_Z^k t_W^{n-k}~. 
\ee
Unrefining with respect to the accidental $U(1)$ symmetry sets $t_Z = t_W$ and yields 
\be
g(T; L^{k, n-k,k})|_{t_Z = t_W} = \PE \left[ t_X + t_Y + 2 t_W - t_W^n\right]~, \qquad \mathrm{with} \quad t_X t_Y = t_W^n~, 
\ee
which indeed coincides with \eqref{unref_orbifold} if $(t_X,t_Y,t_W)=(t_x,t_y,t_w)$.

\section{The string amplitude \label{samp}}

In the last section we give some evidence for the interpretation of mass deformations as complex deformations of the UV Calabi-Yau cone. This suggests that mass deformations can be realized in String Theory by turning on 3-form NSNS and RR fluxes.
Here we support this identification, by computing the mass couplings of 3-form fluxes to open string fermion bilinears.
 The mass couplings will be extracted from the three point functions on a disk involving the insertion of two open and one closed string vertex operators.\footnote{M.~B, J.~F.~M. and D.~R.~P. would like to acknowledge stimulating discussions on this issue with G.~Inverso and L.~Martucci.}

  We focus on the $\mathbb{C}^2/\mathbb{Z}_n \times \mathbb{C}$ case. The generalization to the other orbifold theories under consideration here is straightforward since the orbifold groups  always contain an ${\cal N}=2$ element, i.e. an element leaving invariant one complex plane, let us say $X^3$. Turning on a flux belonging to this sector will give mass to the scalar field  $\Phi^3$. The only difference with the ${\cal N}=2$ setup is that for ${\cal N}=1$ orbifold theories, the orbifold group acts non-trivially on  $\Phi^3$ and therefore Chan-Paton indices should be taken off-diagonal leading to bifundamental rather than adjoint representations.

Open string vertices can be chosen among the gaugino and the fermions in the bifundamental matter 
\bea
V_{\Lambda^0} &=& \Lambda^0_\alpha \, e^{-{\varphi\over 2} } S^\alpha\,  \Sigma_0 \,  (x)   \nn\\
V_{\Lambda^I} &=& \Lambda^I_{ \alpha} \, e^{-{\varphi\over 2}} S^{\alpha}\,  \Sigma_I \,  (x)  ~.
\eea
 The gaugino field is described by block diagonal matrix $\Lambda^0_\alpha$ while matter fermions
   are given by off-diagonal matrices  $\Lambda^I_{\dot \alpha}$  with
 non-trivial $N_a\times N_{a+a_I}$ block components. 
 There are two choices for the closed string fields depending on whether we consider fluxes coming from the untwisted or twisted sectors. The result in the untwisted sector can be borrowed from that in flat space-time \cite{Billo:2008pg} after Chan-Paton matrices are properly projected. For convenience of the reader, we review the results here.    
 The closed string vertices for RR and NSNS 3-form fluxes in the untwisted sector are given by  
 \bea
 V_F&=&   ({\cal F} {\cal R}_0)_{AB}   \, e^{-{\varphi\over 2} } S^\alpha\,  \Sigma_A (z)\, e^{- {\varphi\over 2} } \,S_\alpha\,   \Sigma_B   (\bar z)+  ({\cal F} {\cal R}_0)^{AB}   \, e^{-{\varphi\over 2} } C^{\dot \alpha}\,  \Sigma^A (z)\, e^{- { \varphi\over 2}} \,C_{\dot \alpha}\,   \Sigma^B   (\bar z) \nn\\
 V_H &=& \partial_m (B R_0)_{np}  \, \psi^m \psi^n (z)  e^{-\varphi}  \psi^p  (\bar z)    
 \eea
   with $A=0,\ldots 3$ upper and lower indices labeling the L- and R-moving spinor representations of the R-symmetry group SO(6), and $R_0=-1$, ${\cal R}_0=\Gamma_4\ldots \Gamma_9$ are the reflection matrices relating left to right moving modes of the strings. Explicitly
   \be
  ( {\cal F} {\cal R}_0)^{AB} =*F_{mnp} (\Sigma^{mnp})^{AB}    \qquad   (B R_0)_{np} =-B_{np}
   \ee
where $m=1,..6$ runs over the vector of $SO(6)$. 
   Finally internal spin fields can be written in the bosonized form 
      \be
   \Sigma_0=e^{{i\over 2} (\varphi_1+\varphi_2+\varphi_3)} \qquad  \Sigma_I=e^{-{i\over 2} (\varphi_1+\varphi_2+\varphi_3-2\varphi_I)} \qquad S_\alpha=e^{\pm {i\over 2} (\varphi_4+\varphi_5)}  \qquad 
   C_{\dot\alpha}=e^{\pm {i\over 2} (\varphi_4-\varphi_5)}
   \ee
   and upper indices are given by their complex conjugates. Collecting all pieces and plugging them into the disk amplitudes
   $\langle V_{F,H} \Lambda^A \Lambda^B \rangle $,  one finds  the three point couplings \cite{Billo:2008pg}
 \be
{\cal L}_{\rm 3-form}=\ft{2\pi }{3!}\,   G^{IASD}_{mnp}    \, (\bar{\Sigma}^{mnp})_{AB} \, 
 {\rm Tr} \Lambda^{\alpha A} \,  \Lambda_\alpha^B +{\rm h.c.}  \label{3form}
\ee   
where $G^{IASD}=*F-\tau  H$ is the imaginary anti self-dual part of the three-form field\footnote{We take $\tau={i \over g_s} $. }  
 \be
 G=F-\tau H~.
 \ee
 In components, after keeping only $\mathbb{Z}_n$-invariant components of the fluxes, one finds
 \be
\ft{1}{2\pi} {\cal L}_{\rm untw}=    G_{(3,0)}    
 {\rm Tr} \Lambda^{\alpha 0} \,  \Lambda_\alpha^0 + G_{(1,2)}  {\rm Tr}  \Lambda^{\alpha 3} \,  \Lambda_\alpha^3  ~.
    \label{3form2}  
 \ee
  Notice that the first term breaks supersymmetry since it gives mass to the gaugino. On the other hand, a flux of (1,2) type
  generates a supersymmetric mass for the  adjoint fermions $\Lambda^3$.   
    
Now, let us consider the twisted closed string spectrum. 
Denoting by $\theta_I=(\theta,1-\theta,0) $,  $\theta\in \ft{1}{n} \mathbb{Z}$, the bosonic twists, the relevant vertex for NSNS and RR field strengths can be written as
\bea
V_{H,{\rm tw}} &=&   H\,    e^{-\varphi}    \prod_{I=1}^2 \sigma_{\theta_I} e^{i \varphi_I \theta_I }    (z)
e^{-i \varphi_3} \, \prod_{I=1}^2 \sigma_{\theta_I} e^{-i \varphi_I \theta_I }   (\bar z) \nn\\
V_{F,{\rm tw}} &=&   F\,    e^{-{\varphi\over 2} }  \,e^{-{i \varphi_3\over 2} } \,  S^{\alpha}\,  \prod_{I=1}^2 \sigma_{\theta_I} e^{i \varphi_I({1\over 2}-\theta_I) }    (z)
e^{-{\varphi\over 2} }   \,e^{-{i \varphi_3\over 2}} \,S_{\alpha}  \,  \prod_{I=1}^2 \sigma_{\theta_I} e^{-i \varphi_I({1\over 2} -\theta_I) }   (\bar z)  \label{complexdef}
\eea  
with   $\sigma_{\theta_I}$  the bosonic twist fields.
The vertices \eref{complexdef} are massless for any choice of $\theta$ since left and right moving conformal dimensions add up to one.\footnote{The dimensions of the various fields are
$[e^{-\varphi/2}]=\ft38$, $[e^{-\varphi}]=\ft12$, $[e^{i q \varphi_i} ]=\ft{q^2}{2}$, $[\sigma_{\theta_I}]=   \ft{\theta_I}{2}(1-\theta_I)$, $[e^{i \theta}]=\ft{\theta^2}{2}$. }
 Terms combine again into imaginary anti-self-dual combinations generating the bilinear couplings
\be
\ft{1}{2\pi} {\cal L}_{\Theta^h-{\rm tw}}=   G_{(1,2),h}   {\rm Tr} \left( \Theta^h 
\Lambda^{\alpha 3} \,  \Lambda_\alpha^3 \right) ~.  
   \label{3form3}  
 \ee
Comparing \eref{3form3} with \eref{3form2}, one notices that an untwisted 3-form flux produces identical masses for all the adjoint hypermultiplets while fluxes from the twisted sector can be used to tune mass differences. The flows studied in this paper are then induced by NSNS/RR 3-form fluxes coming from twisted sectors localized at the singularities.

 In the following we present the derivation of this coupling for the RR vertex. A similar computation can be performed for the NSNS field.
\\

\subsection{The RR amplitude}

Let us compute the coupling $F \Lambda^{\alpha 3} \Lambda_\alpha^3$.    The relevant disk amplitude (at zero momenta) is
\be
\int dz_2 \, \langle V_{\Lambda^3} (z_1) V_{\Lambda^3}(z_2)  V_F  (z_3,z_4) \rangle  = F \Lambda^3_{\alpha_1} \Lambda^3_{\alpha_2} \epsilon_{\alpha_3 \alpha_4} \, {\cal A}^{\alpha_1 \alpha_2 \alpha_3 \alpha_4}
\label{afm}
\ee
with
\bea
&& {\cal A}^{\alpha_1 \alpha_2 \alpha_3 \alpha_4} =\int dz_2 ~ \bigg\langle 
c e^{-{\varphi\over 2}} S^{\alpha_1}\,  \Sigma_3 \,  (z_1)  e^{-{\varphi\over 2} } S^{\alpha_2}\,  \Sigma_3 \,  (z_2)  \nn\\
&&~~~~ \times 
 c e^{-{\varphi\over 2} }  \,e^{- i \varphi_3\over 2} \,  S^{\alpha_3}\,  \prod_{I=1}^2 \sigma_{\theta_I} e^{i \varphi_I({1\over 2}-\theta_I) }    (z_3)
c e^{-{ \varphi\over 2} }   \,e^{- i \varphi_3\over 2} \,S^{\alpha_4}  \,  \prod_{I=1}^2 \sigma_{\theta_I} e^{-i \varphi_I({1\over 2} -\theta_I) }   (z_4) 
 \bigg\rangle~.\nn
\eea
 The various contributions are
\bea
&& \langle e^{-{\varphi\over 2} }(z_1) e^{-{\varphi\over 2} } (z_2)  e^{-{\varphi\over 2} }(z_3) e^{-{\varphi\over 2} }(z_4)  \rangle =(z_{12} z_{13} z_{14} z_{23}z_{24} z_{34} )^{-1/4} \nn\\
&& \langle e^{ {i \varphi_3\over 2}  }(z_1) e^{ {i \varphi_3\over 2} } (z_2)  e^{ -{i \varphi_3\over 2}  }(z_3) e^{ -{i \varphi_3\over 2}   }(z_4)  \rangle =\left({z_{12} z_{34} 
\over z_{13} z_{24}z_{14} z_{23} } \right)^{1/4}  \nn \\
&&  \langle     e^{ {i\over 2} (\varphi_1+\varphi_2)   }    (z_1)      e^{ {i\over 2} (\varphi_1+\varphi_2)   }    (z_2)   \prod_{I=1}^2 
\sigma_{\theta_I} e^{i \varphi_I({1\over 2}-\theta_I) }    (z_3)
 \prod_{I=1}^2    \sigma_{\theta_I} e^{-i \varphi_I({1\over 2} -\theta_I) }   (z_4)
  \rangle = \left( {1\over z_{12} z_{34}}\right)^{{1\over 2}} \nn \\
&& \langle S^{\alpha_1}(z_1) S^{\alpha_2}(z_2)  S^{\alpha_3}(z_3) {S}^{\alpha_4}(z_4)  \rangle =  \left(  z_{14} z_{23} z_{13} z_{24}   \over z_{12} z_{34} \right)^{1/2} 
\, \left[ {\epsilon^{\alpha_1 \alpha_3} \epsilon^{\alpha_2 \alpha_4}   \over z_{13}z_{24} } + {\epsilon^{\alpha_1 \alpha_4} \epsilon^{\alpha_2 \alpha_3}   \over z_{14}z_{23} } \right]
 \nn\\
&&    \langle c (z_1) c(z_3) c(z_4) \rangle =  z_{13} z_{34} z_{41}~.
\eea
Contracting with $\epsilon_{ \alpha_3 \alpha_4}$, one finds\footnote{ Here we write $ {dz_2 z_{34} \over z_{23} z_{24} }=  {dw\over w}$ with $|w| =1$. } 
\be 
\epsilon_{ \alpha_3 \alpha_4} {\cal A}^{\alpha_1 \alpha_2 \alpha_3 \alpha_4}  = \epsilon^{\alpha_1 \alpha_2}  \, \int_{|w|=1}{ dw \over w}  =2\pi i \, \epsilon^{\alpha_1 \alpha_2} 
\label{resa} \ee 
 with $w={z_{24} z_{13} \over z_{14} z_{23}} $ the complex cross ratio.   Plugging \eref{resa} into \eref{afm} one finds the F-contribution to the coupling \eref{3form3}.  

A similar computation can be performed for untwisted R-R fluxes. 
   The relevant disk amplitude (at zero momenta) reads
  \bea
 {\cal A}_{IJ}^{KL}  &=& \int dz_2  \left\langle 
c e^{-{\varphi\over 2}} S^{\alpha }\,  \Sigma_I \,  (z_1)  e^{-{\varphi\over 2} } S_{\alpha }\,  \Sigma_J \,  (z_2)  
 c e^{-{\varphi\over 2} }  \,e^{- i \varphi_3\over 2} \,  C^{\dot \alpha }\,     \Sigma^K \,    (z_3)
c e^{-{ \varphi\over 2} }   \,e^{- i \varphi_3\over 2} \,C_{\dot \alpha }\,     \Sigma^L \,    (z_4)  \right\rangle
\nn\\
&=& \delta^{(I}_{(K}  \delta^{J)}_{L)} \, \int_{|w|=1}{ dw \over w}  =2\pi i \,\delta^{(I}_{(K}  \delta^{J)}_{L)}
\eea 
leading to \eref{3form2}. 
\\

\section{Conclusions}\label{sec:conclusions}

In this paper we have shown that brane tiling methods can be used to efficiently study RG flows between toric quiver gauge theories triggered by mass terms. Even though mass terms break the $U(1)^3$ toric symmetry of the UV superconformal field theory and cannot be described by nodes in a brane tiling, for judicious choices of the masses with pairs of equal and opposite mass parameters it is possible to flow to another superconformal toric quiver gauge theory in the IR. The IR toric $U(1)^3$ non-baryonic symmetry involves an accidental mesonic symmetry which appears once the massive fields are integrated out. The accidental symmetry is manifested by a change of variables for the light fields which recasts the superpotential in toric form. 
The endpoints of such renormalization group flows can be easily visualized by performing a certain move on the brane tiling associated to the UV fixed point.  The effect of this move is to reverse the winding numbers of a particular zig-zag path made of those massless fields which appear in two toric superpotential terms with massive fields. 

Although the mass-deformed theory does not have a $U(1)^3$ non-baryonic symmetry along the RG flow, its mesonic moduli space is actually a toric Calabi-Yau cone: this is nothing but the Calabi-Yau cone associated to the IR fixed point, since the U(1) charges of $F$-terms are independent of the RG scale.  We have shown that the volumes of the Sasaki-Einstein bases of the singularity cones always increase along the flows from UV to IR, in agreement with the holographic $a$-theorem. Interestingly, for flows starting from an orbifold singularity  the ratio between IR and UV volumes depends only on the order of the group and the number of massive deformations with maximum value $32/27$ matching the universal value explained in \cite{Tachikawa:2009tt} for flows from $\mathcal{N}=2$ to  $\mathcal{N}=1$ SCFT where all adjoint fields get masses. This universal value is also achieved for flows involving massive bifundamental matter.
  
The toric Calabi-Yau cones associated to the UV and IR fixed points are related by a complex deformation. We have shown the relation between UV and IR toric Calabi-Yau cones in a simple class of examples, in terms of the algebraic description of the singularity and its Hilbert series. The introduction of a mass term has the net effect of partially unrefining the Hilbert series of the mesonic moduli space of the UV theory with respect to the global symmetry broken by the mass term, and the Hilbert series of the mass-deformed UV theory coincides with that of the Hilbert series of the IR theory, unrefined with respect to the accidental symmetry. 

The mass deformation is induced by the presence of an imaginary-self-dual 3-form fluxes in the twisted sector. We supported this identification by an explicit computation of the disk amplitude involving a closed 3-form vertex  from the twisted sector and two open string fermions.

Analogously to our study of mass deformations, it would be interesting to analyze the effect of other relevant or marginal deformations \cite{Butti:2006nk} and identify the source of the deformation from the bulk point of view. 
The results of this paper relate gauge theories on non-orbifold singularities like $L^{a b a}$ (or orbifolds of them)  to orbifold theories.
In orbifold theories, one has complete control of the dynamics from the world-sheet vantage point. One can not only compute mass terms generated by twisted bulk fluxes, as done in section \sref{samp}, but also superpotentials and other interactions in the effective action dynamically generated by `gauge' or `exotic' instantons \cite{Ibanez:2006da,Florea:2006si,Cvetic:2007ku,Argurio:2007vqa,Bianchi:2007wy,Bianchi:2007fx,Bianchi:2009bg}.  It would be interesting to extend the results of this paper to superconformal unoriented theories that emerge from D3-branes at orientifold singularities \cite{Bianchi:2013gka} and exploit the worldsheet description in the UV to learn about the strong coupling dynamics of the non-orbifold theories in the IR. Given their possible role in embedding (supersymmetric) extensions of the Standard Model, configurations of unoriented D-branes at Calabi-Yau singularities certainly deserve a systematic analysis starting from the toric case along the lines of \cite{ Blumenhagen:2005mu,Blumenhagen:2006ci,Bianchi:2009ij,Ibanez:2012zz}.

\section*{Acknowledgements}

We kindly acknowledge discussions with A.~Amariti, C.~Bachas, M.~Bertolini,  
S.~Franco,  A.~Mariotti, D.~Orlando, G.~Inverso and L.~Martucci.
We thank the following institutes for their kind hospitality during the completion of this work: the Theoretical Physics groups at Imperial College London (MB, JFM, DRP, R-KS), Queen Mary University of London (MB), Padua University (R-KS), the University of Rome Tor Vergata (SC, DRP) and the Mathematical Institute of the University of Oxford (JFM); the Simons Center of Geometry and Physics in Stony Brook (AH, R-KS), the Galileo Galilei Institute in Florence (SC), the Royal Society at Chicheley Hall (MB, SC, AH, R-KS) and the ICMS in Edinburgh (MB, SC, AH, JFM, R-KS). The work of MB and JFM is partially supported by the ERC Advanced Grant n. 226455 ``Superfields" and was initiated while MB was visiting Imperial College and largely carried on while MB was holding a Leverhulme Visiting Professorship at QMUL. The work of JFM is also supported by the Engineering and Physical Sciences Research Council, grant numbers  EP/I01893X/1 and EP/K034456/1. The research of DRP was supported by the Padova University Project CPDA119349 and by the MIUR-PRIN contract 2009-KHZKRX.

\begin{appendix}

\section{Details of the flows \label{sappendix}}

In this appendix we collect the details of the mass flows  described in section \sref{s1n1}.  For each case we list the starting and end superpotentials, the mass deformation, the F-term conditions and the field redefinitions. 
\\

\subsection{$\mathbb{C}^2/\mathbb{Z}_n\times\mathbb{C}$  to $L^{k,n-k,k}$\label{Lknk}}

 Superpotentials and mass terms:  
\beal{es100n1}
W_{\mathbb{C}^2/\mathbb{Z}_n\times\mathbb{C}}&=& \sum_{i=1}^{n} \phi_i \left( X_{i,i-1} X_{i-1,i} -X_{i,i+1} X_{i+1,i} \right) \nn\\
\Delta W &=& \frac{m}{2}\sum_{i=1}^k\left(\phi_{2i-1}^2 - \phi_{2i}^2 \right)\\
 W_{L^{k,n-k,k}}&=&\sum_{i=1}^{k} \left(X_{2i-1,2i}^\prime X_{2i,2i-1}^\prime X_{2i-1,2i-2}X_{2i-2,2i-1}-X_{2i,2i-1}^\prime X_{2i-1,2i}^\prime X_{2i,2i+1}X_{2i+1,2i}\right)\nn\\
&+&\sum_{i=k+1}^{n} \phi_i^\prime\left(X_{i,i-1}X_{i-1,i}-X_{i,i+1}X_{i+1,i}\right) \nn
\eea
where subscripts  $i$ are understood modulo  $n$.  
  
F-term and field redefinitions:
\beal{fshiftk}
\phi_{2i-1} &=& \frac{1}{m} \left( X_{2i-1,2i} X_{2i,2i-1} - X_{2i-1,2i-2} X_{2i-2,2i-1} \right) ~~\nn\\
\phi_{2i}& =& \frac{1}{m} \left( X_{2i,2i-1} X_{2i-1,2i} - X_{2i,2i+1} X_{2i+1,2i} \right)  ~~\nn\\
\phi_j &=& \phi_j^\prime - \frac{1}{2m}\left( X_{j,j+1} X_{j+1,j} + X_{j,j-1} X_{j-1,j}\right) ~~\nn\\
X_{2i-1,2i} X_{2i,2i-1} &=& m X_{2i-1,2i}^\prime X_{2i,2i-1}^\prime ~~
\eea
with $i=1,\ldots,k$ and $j=2k+1,\ldots,n$.
\\
 
\subsection{$(\mathbb{C}^2/\mathbb{Z}_n\times \mathbb{C})/\mathbb{Z}_2$ to $L^{k,n-k,k}/\mathbb{Z}_2$}

The superpotential of the $(\mathbb{C}^2/\mathbb{Z}_{n}\times \mathbb{C})/\mathbb{Z}_{2}$  model is given by
\beal{Wc2ZnZ2}
W_{(\mathbb{C}^2/\mathbb{Z}_n\times \mathbb{C})/\mathbb{Z}_2}&=&
\sum_{i=1}^{n} \Big[ X_{2i-1,2i} \left(X_{2i,2i+2} X_{2i+2,2i-1} - X_{2i,2i-3} X_{2i-3,2i-1}\right)
\nn\\
&&
\hspace{1cm}
+
X_{2i,2i-1} \left(X_{2i-1,2i+1} X_{2i+1,2i} - X_{2i-1,2i-2} X_{2i-2,2i}\right)
\Big]\nn\\
\Delta W &=&m \sum_{i=1}^{2k} (-1)^{i}\,X_{2i-1,2i} X_{2i,2i-1} \\
W_{L^{k,n-k,k}/\mathbb{Z}_2}&=&
\sum_{i=2k+1}^{n} \Big[ X'_{2i-1,2i} \left(X_{2i,2i+2} X_{2i+2,2i-1} - X_{2i,2i-3} X_{2i-3,2i-1}\right)
\nn\\
&+&
X'_{2i,2i-1} \left(X_{2i-1,2i+1} X_{2i+1,2i} - X_{2i-1,2i-2} X_{2i-2,2i}\right)
\Big]~\nn\\
&+& \sum_{i=1}^k X'_{4i-3,4i-1} \left(X_{4i-1,4i+1} X_{4i+1,4i} X_{4i,4i-3} - X_{4i-1,4i-2} X_{4i-2,4i-5}X_{4i-5,4i-3}\right)\nn\\
&+& \sum_{i=1}^k X'_{4i-2,4i} \left(X_{4i,4i+2} X_{4i+2,4i-1} X_{4i-1,4i-2} - X_{4i,4i-3} X_{4i-3,4i-4}X_{4i-4,4i-2}\right)\nn
\eea
where subscripts  $i$ are now understood modulo  $2n$. \\

F-terms and field redefinitions:
\beal{FtermshiftsZnZ2gen}
X_{2i-1,2i} &=& (-1)^{i+1} \frac{1}{m}\left(X_{2i-1,2i+1}X_{2i+1,2i}-X_{2i-1,2i-2} X_{2i-2,2i}\right) \nn\\
X_{2i,2i-1}&=& (-1)^{i+1}\frac{1}{m}\left(X_{2i,2i+2}X_{2i+2,2i-1}-X_{2i,2i-3} X_{2i-3,2i-1}\right) \nn\\
X_{2j-1,2j} &=& X'_{2j-1,2j} -\frac{1}{2m} ( X_{2j-1,2j-2} X_{2j-2,2j}+ X_{2j-1,2j+1} X_{2j+1,2j})
\nn\\
X_{2j,2j-1} &=& X'_{2j,2j-1} -\frac{1}{2m} ( X_{2j,2j+2} X_{2j+2,2i-1} + X_{2j,2j-3} X_{2j-3,2j-1}) \nn\\
X_{4l-3,4l-1} &=& m X'_{4l-3,4l-1} ~,~ \qquad\qquad\qquad X_{4l-2,4l} = m X'_{4l-2,4l} ~,~
\eea
with $i=1,\ldots,2k$, $j=2k+1,\dots,n$ and $l=1,\dots,k$.

\subsection{ $\mathbb{C}^3/\mathbb{Z}_{2n}$ to  $L^{k,n-k,k}/ \mathbb{Z}'_{2}$ \label{sapp3}}

 Superpotentials and mass terms:  
\beal{Wc3z2nbis}
W_{  \mathbb{C}^3/\mathbb{Z}_{2n} }&=&
\sum_{i=1}^{2n} X_{i,i+1} \left(X_{i+1,i+n+1} X_{i+n+1,i} - X_{i+1,i+n} X_{i+n,i}\right)\nn\\
\Delta W &=&
 m \sum_{i=1}^k \left( X_{2i-1,2i-1+n} X_{2i-1+n,2i-1} - X_{2i,2i+n} X_{2i+n,2i} \right) \\
 W_{L_{k,n-k,k}/ \mathbb{Z}'_{2}}&=&
\sum_{i=2k+1}^{n}  X'_{i,i+n} \left(X_{i+n,i-1} X_{i-1,i} - X_{i+n,i+n+1}X_{i+n+1,i} \right) + \nn \\
&+& \sum_{i=2k+1}^{n} X'_{i+n,i} \left( X_{i,i+n-1} X_{i+n-1,i+n}- X_{i,i+1} X_{i+1,i+n} \right)+\nn\\
&+&\sum_{i=1}^k X'_{2i-1,2i}\left( X_{2i,2i-1+n} X_{2i-1+n,2i-2} X_{2i-2,2i-1} - X_{2i,2i+1}X_{2i+1,2i+n}X_{2i+n,2i-1} \right)+ \nn\\
&+&\sum_{i=1}^k X'_{2i-1+n,2i+n} \big( X_{2i+n,2i-1} X_{2i-1,2i-2+n} X_{2i-2+n,2i-1+n}+ \nn \\
& & \qquad\qquad\qquad~ -X_{2i+n,2i+1+n}X_{2i+1+n,2i} X_{2i,2i-1+n}\big)\nn\\
\eea
with the subscripts understood modulo $2n$.
 
F-terms and field redefinitions:
\beal{c3z2nfshitsgen}
X_{2i-1,2i-1+n} &=& \frac{1}{m}\left( X_{2i-1,2i}X_{2i,2i-1+n}-X_{2i-1,2i-2+n} X_{2i-2+n,2i-1+n} \right) \nn\\
X_{2i-1+n,2i-1} &=& \frac{1}{m}\left( X_{2i-1+n,2i+n} X_{2i+n,2i-1}-X_{2i-1+n,2i-2} X_{2i-2,2i-1} \right)  \nn\\
X_{2i,2i+n} &= & \frac{1}{m}\left( X_{2i,2i+n-1} X_{2i+n-1,2i+n}-X_{2i,2i+1} X_{2i+1,2i+n} \right) \nn\\
X_{2i+n,2i} &=& \frac{1}{m}\left( X_{2i+n,2i-1} X_{2i-1,2i}-X_{2i+n,2i+1+n} X_{2i+1+n,2i} \right) \nn\\
X_{j,j+n} &=& X'_{j,j+n} -\frac{1}{2m} ( X_{j,j+1} X_{j+1,j+n}+ X_{j,j+n-1} X_{j+n-1,j+n}) \nn\\
X_{j+n,j} &=& X'_{j+n,j} -\frac{1}{2m} (X_{j+n,j+n+1} X_{j+n+1,j} + X_{j+n,j-1} X_{j-1,j}) \nn\\
X_{2i-1,2i} &=& m X'_{2i-1,2i} ~,~ \quad\quad\quad X_{2i-1+n,2i+n} = m X'_{2i-1+n,2i+n} ~,~
\eea
for $i=1,\ldots,k$ and $j=2k+1,\ldots,n$.
\\

\subsection{$\mathrm{PdP}_{4b} $ to $ \mathrm{PdP}_{4a} $   \label{s_ex2} \label{sapp4}}
 
Superpotentials and mass terms:  \beal{es1}
W_{  \mathrm{PdP}_{4b}   } &=& 
X_{1\hspace{0.04cm}2}X_{2\hspace{0.04cm}5}X_{5\hspace{0.04cm}1}+
X_{1\hspace{0.04cm}3}X_{3\hspace{0.04cm}4}X_{4\hspace{0.04cm}1}+X_{1\hspace{0.04cm}4}X_{4\hspace{0.04cm}7}X_{7\hspace{0.04cm}1}+X_{2\hspace{0.04cm}4}X_{4\hspace{0.04cm}5}X_{5\hspace{0.04cm}2}+X_{3\hspace{0.04cm}5}X_{5\hspace{0.04cm}6}X_{6\hspace{0.04cm}3}\nn\\
&-&X_{1\hspace{0.04cm}2}X_{2\hspace{0.04cm}4}X_{4\hspace{0.04cm}1}-X_{1\hspace{0.04cm}3}X_{3\hspace{0.04cm}7}X_{7\hspace{0.04cm}1}-X_{1\hspace{0.04cm}4}X_{4\hspace{0.04cm}5}X_{5\hspace{0.04cm}1}-X_{2\hspace{0.04cm}3}X_{3\hspace{0.04cm}5}X_{5\hspace{0.04cm}2}-X_{2\hspace{0.04cm}5}X_{5\hspace{0.04cm}6}X_{6\hspace{0.04cm}2}\nn\\
&+&X_{2\hspace{0.04cm}3}X_{3\hspace{0.04cm}7}X_{7\hspace{0.04cm}6}X_{6\hspace{0.04cm}2}-X_{3\hspace{0.04cm}4}X_{4\hspace{0.04cm}7}X_{7\hspace{0.04cm}6}X_{6\hspace{0.04cm}3}
\nn\\
\Delta W &=&
 m \,(X_{1\hspace{0.04cm}4} X_{4\hspace{0.04cm}1} - X_{2\hspace{0.04cm}5} X_{5\hspace{0.04cm}2}) \nn\\
 W_{ \mathrm{PdP}_{4a}  }
&=& X'_{4\hspace{0.04cm}5} \left(X_{5\hspace{0.04cm}1} X_{1\hspace{0.04cm}3} X_{3\hspace{0.04cm}4} - X_{5\hspace{0.04cm}6}X_{6\hspace{0.04cm}2}X_{2\hspace{0.04cm}4}\right) +
X'_{1\hspace{0.04cm}2} \left( X_{2\hspace{0.04cm}4} X_{4\hspace{0.04cm}7}X_{7\hspace{0.04cm}1}-
X_{2\hspace{0.04cm}3} X_{3\hspace{0.04cm}5}X_{5\hspace{0.04cm}1}\right)+
\nn\\
&+& X'_{6\hspace{0.04cm}3} \left( X_{3\hspace{0.04cm}5} X_{5\hspace{0.04cm}6} - X_{3\hspace{0.04cm}4} X_{4\hspace{0.04cm}7} X_{7\hspace{0.04cm}6}\right) + X'_{3\hspace{0.04cm}7} \left(X_{7\hspace{0.04cm}6} X_{6\hspace{0.04cm}2} X_{2\hspace{0.04cm}3}-  X_{7\hspace{0.04cm}1} X_{1\hspace{0.04cm}3} \right)~.
\eea
 F-terms and field redefinitions:
\beal{es3}
X_{1\hspace{0.04cm}4} &=& + \frac{1}{m} (X_{1\hspace{0.04cm}2} X_{2\hspace{0.04cm}4} - X_{1\hspace{0.04cm}3} X_{3\hspace{0.04cm}4}) ~,~\quad
X_{4\hspace{0.04cm}1} = + \frac{1}{m} (X_{4\hspace{0.04cm}5} X_{5\hspace{0.04cm}1} - X_{4\hspace{0.04cm}7} X_{7\hspace{0.04cm}1}) ~,~
\nn\\
X_{2\hspace{0.04cm}5} &=&  \frac{1}{m} (X_{2\hspace{0.04cm}4} X_{4\hspace{0.04cm}5} - X_{2\hspace{0.04cm}3} X_{3\hspace{0.04cm}5}) ~,~\quad 
X_{5\hspace{0.04cm}2} = - \frac{1}{m} (X_{5\hspace{0.04cm}1} X_{1\hspace{0.04cm}2} - X_{5\hspace{0.04cm}6} X_{6\hspace{0.04cm}2}) ~,~
 \\
 X_{3\hspace{0.04cm}7} &=& X'_{3\hspace{0.04cm}7} - X_{3\hspace{0.04cm}4} X_{4\hspace{0.04cm}7}~,
\quad
X_{6\hspace{0.04cm}3} = X'_{6\hspace{0.04cm}3} - X_{6\hspace{0.04cm}2} X_{2\hspace{0.04cm}3}~, \quad
X_{1\hspace{0.04cm}2} = m X'_{1\hspace{0.04cm}2} ~,\quad
X_{4\hspace{0.04cm}5} = m X'_{4\hspace{0.04cm}5} ~. \nn
\eea

\end{appendix}

\bibliographystyle{JHEP}
\bibliography{mybib}

\end{document}